\documentclass{article}
\usepackage{jheppub}
\usepackage{physics}
\usepackage{tikz}
\usepackage{comment}
\usepackage{caption}
\usepackage{subcaption}

\newcommand\titlemath[1]{\texorpdfstring{#1}{Lg}}

\usepackage{xcolor}

\begin{document}

\title{A spin on Hagedorn temperatures and string stars}

\author[a]{Josef Seitz}
\author[b]{and Erez Y.~Urbach}
\affiliation[a]{Department of Particle Physics and Astrophysics, Weizmann Institute of Science, Rehovot, Israel}
\affiliation[b]{School of Natural Sciences, Institute for Advanced Study, Princeton, NJ 08540, USA}
\emailAdd{josef-emanuel.seitz@weizmann.ac.il}
\emailAdd{urbach@ias.edu}

\abstract{
We discuss the correspondence between highly excited strings and black holes in the presence of angular momentum. At fixed imaginary angular velocity $\nu$, we show that free strings exhibit a Hagedorn instability due to a thermal-winding mode turning tachyonic. This allows us to determine the exact Hagedorn temperature $\beta_H(\nu)$ for bosonic, type II, and heterotic strings. 
Using the effective field theory for the thermal-winding mode around the $\nu$-dependent background, we find a novel `rotating string star' saddle (perturbatively in the angular velocity) and study its properties.
This configuration describes a self-gravitating bound state of highly excited rotating strings.
As in the non-rotating case, the saddle is qualitatively shown to interpolate between the rotating strings phase and a rotating black hole. 
We also comment on the implications of these results for anti-de Sitter space.
}
\maketitle

\section{Introduction}
Over time, black holes tend to shrink as they radiate.
The fate of small (highly-curved) black holes in a UV-complete theory of gravity is therefore an interesting self-consistency check~\cite{Bedroya:2022twb,Basile:2023blg,Bedroya:2024ubj,Bedroya:2024uva,Herraez:2024kux,Herraez:2025gcf,Herraez:2025clp}, which could teach us a lesson on black hole microscopy. In string theory, the nature of black holes as their horizon size reaches the string scale remains an open question~\cite{Susskind:1993ws,Maldacena:1996ky,Horowitz:1996nw,Peet:2000hn}.
One version of this question is phrased in terms of Euclidean gravity~\cite{Chen:2021dsw}. The Euclidean black hole saddle in asymptotically $\mathbb R^d \times S^1_\beta$ string theory\footnote{At finite $g_s$, it is impossible to fix the asymptotic length $\beta$~\cite{Atick:1988si}. We work at arbitrarily small $g_s \ll 1$ throughout the paper, in which the problem disappears. Another option is to embed the discussion inside thermal AdS (see~\cite{Urbach:2022xzw} and section \ref{sec:ads}).} is well described by Einstein gravity for $\beta \gg l_s$, but what is its description for $\beta \sim l_s$?
Horowitz and Polchinski~\cite{Horowitz:1997jc} suggested that at high temperatures and for $3\le d \le 5$, the saddle is a normalizable condensate of strings winding on the thermal cycle of $\mathbb R^d \times S^1_\beta$. The saddle, interpreted as a self-gravitating stringy condensate~\cite{Damour:1999aw,Khuri:1999ez}, is termed `the Horowitz-Polchinski solution' or a `string star'. Surprisingly, this saddle carries an $O(G_N^{-1})$ thermal entropy, and qualitatively agrees with black hole thermodynamics at lower temperatures.\footnote{In~\cite{Chen:2021dsw} a worldsheet index argument was given against a smooth transition between the two descriptions in type II string theory. To the best of our understanding, the argument carries over to our setting ($\nu \ne 0$) as well.}
Recently, there has been a renewed interest in string stars, extending the original setting~\cite{Chen:2021dsw,Mazel:2024alu} to higher dimensions $d>5$~\cite{Balthazar:2022szl,Balthazar:2022hno,Bedroya:2024igb,Chu:2025boe,Chu:2025kzl}, and other background geometries~\cite{Urbach:2022xzw,Urbach:2023npi, Halder:2023nlp,Agia:2023skp,
Ishibashi:2025qwn,
Chu:2024ggi,Emparan:2024mbp,Chu:2025fko}.

\begin{figure}
    \centering
    \includegraphics[width=0.5\linewidth]{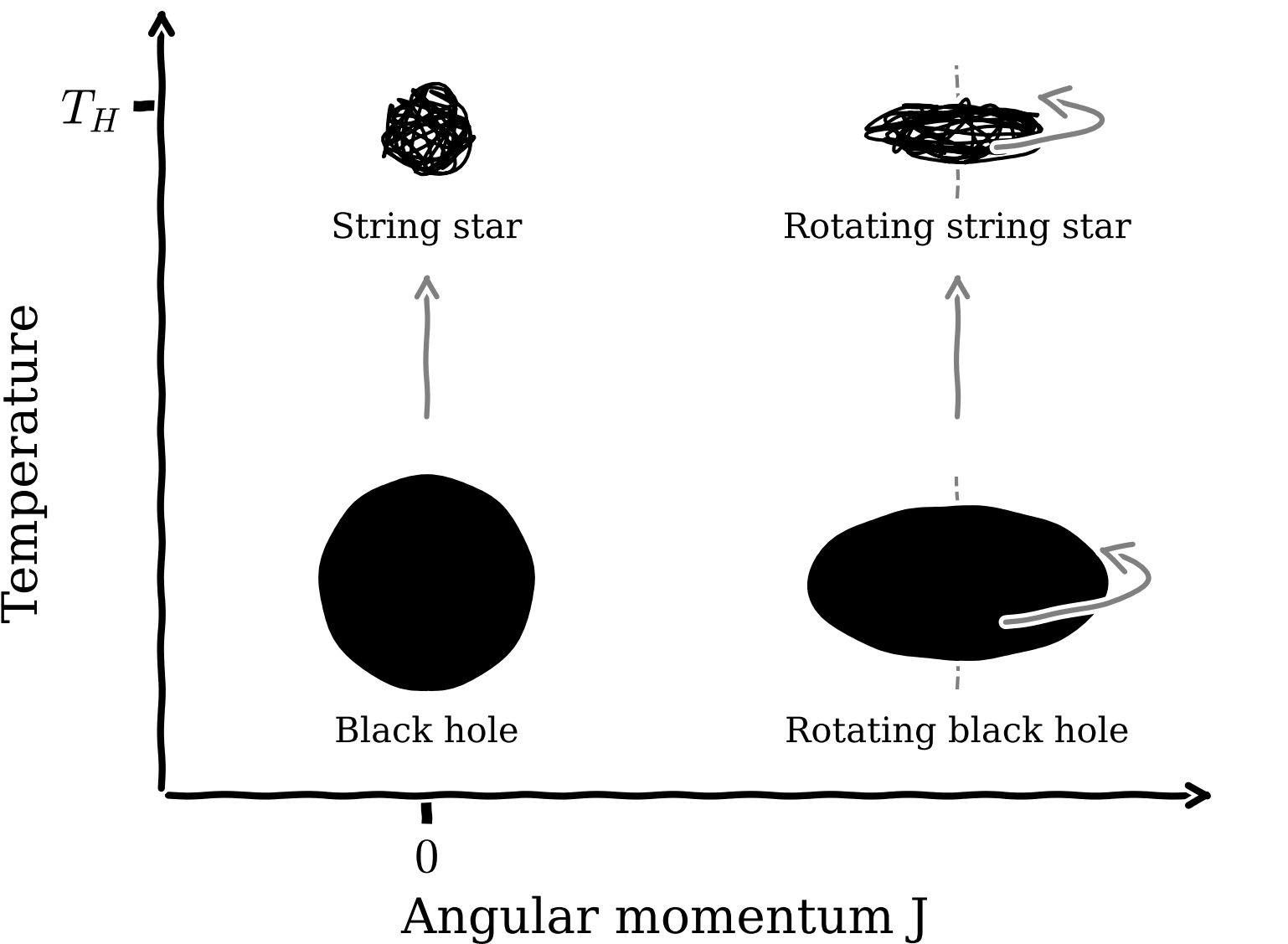}
    \caption{A schematic summary of the suggested black hole/string star correspondence. The Schwarzschild black hole (left) shrinks as the temperature (vertical axis) increases, and turns into a string star close to the string theory Hagedorn temperature $T\lesssim T_H$. Turning on angular momentum (horizontal axis), we find a `rotated' version of the correspondence (right): the rotated black hole turns into a rotating string star at higher temperatures.}
    \label{fig:schem_intro}
\end{figure}

To gather more evidence for this proposal, it is natural to consider extra parameters to turn on, both at the string star and the black hole sides of the correspondence.
In this work, we study the correspondence in the presence of angular momenta.\footnote{Rotating black holes at low temperatures (near-extremal limit) have enjoyed renewed interest recently~\cite{Sen:2012cj,Moitra:2019bub,Kapec:2023ruw,Rakic:2023vhv,Maulik:2024dwq} due to their strong quantum effects at the near-horizon~\cite{Bardeen:1999px} region.}
Let $J$ be a rotation generator in a two-dimensional plane of $\mathbb{R}^d$.
To add angular momenta, we consider an ensemble of fixed temperature $\beta$, and fixed imaginary `angular velocity' potential $\nu$, conjugate to $J$, with the partition function
\begin{equation}\label{eq:beta_nu_ens}
    Z(\beta,\nu) = \sum_{E,J}\exp\left(-\beta E + 2\pi i \nu J\right).
\end{equation}
In terms of Euclidean string theory, this ensemble is defined by an $\mathbb{R}^d \times S^1$ asymptotic boundary condition with a twist $\nu$ between the thermal and the angular directions. The twisted $\mathbb{R}^d\times S^1$ is one possible saddle that contributes to $Z(\beta,\nu)$, describing, at $g_s=0$, freely rotating strings.
The Euclidean rotating black hole is another saddle with the same asymptotics that contributes to \eqref{eq:beta_nu_ens}. This work suggests a description of the rotating black hole saddle when it reaches a string size $\beta \sim l_s$. 
Following the non-rotating case, we find a normalizable winding condensate solution on top of the twisted $\mathbb R^d \times S^1$ background. The result is a `rotating string star' saddle, interpreted as a bound state of self-gravitating rotating strings. By studying the saddle thermodynamic properties, we find a qualitative agreement with the rotating black hole (see figure \ref{fig:schem_intro}).

We begin in section \ref{sec:flat_space_therm} by reviewing the thermodynamic properties of highly-excited rotating strings.\footnote{
For previous literature on this subject,  see~\cite{Russo:1994ev,Matsuo:2009sx}.
Recently~\cite{Ceplak:2023afb,Ceplak:2024dxm} studied aspects of the correspondence between free strings and black holes with angular momenta. While some of our work overlaps with theirs, there are important differences. Most importantly, they compared black holes with free strings, whereas we compare them to the string star. Secondly, all the objects we consider in this work (strings, string stars, or black holes) will possess a `small' amount of angular momentum, and will remain away from extremality.
More abstractly, the process that interpolates between strings and black holes in~\cite{Ceplak:2023afb,Ceplak:2024dxm} was an adiabatic (fixed entropy) process of varying the string coupling. We follow~\cite{Chen:2021dsw}, and change the temperature (and possibly the angular velocity) at fixed string coupling (fixed theory). We believe this difference is not consequential.}
By studying free strings in the $\beta,\nu$ ensemble \eqref{eq:beta_nu_ens}, we find it shares an important property with the canonical (non-rotating $\nu=0$) ensemble: it has a Hagedorn instability! 
Explicitly, we find that due to the high energy spectrum of free rotating strings, $Z(\beta,\nu)$ diverges at temperatures above a $\nu$-dependent Hagedorn temperature $\beta < \beta_H$.\footnote{While we were not aware of this work until the final stages of writing, ~\cite{Mertens:2014nca} is the only previous work we found that presents $\beta_H(\nu)$, in the context of the pure NS-NS bosonic AdS$_3$ background.} Defining $R_H = \beta_H / (2\pi)$, we find (for $|\nu|<1$)
\begin{equation}\label{eq:hag_flat_final}
\begin{split}
    R_H/l_s &=
    \begin{cases}
        \sqrt{4-2|\nu|+2\nu^2}, & \text{Bosonic}\\
        \sqrt{2-2|\nu|}, & \text{Type II}\\
        \frac{1}{2}\left(\sqrt{4-2|\nu|+2\nu^2}+\sqrt{2-2|\nu|}\right),& \text{Heterotic}
    \end{cases}.
\end{split}
\end{equation}
We give two alternative ways to find \eqref{eq:hag_flat_final}. One is to directly check the Hagedorn growth of the partition function due to high-energy excitations. A second, more instructive way for our purposes is to follow~\cite{Sathiapalan:1986db,Kogan:1987jd,Atick:1988si} and interpret the Hagedorn instability as a tachyonic instability of a thermal winding mode in $\mathbb{R}^d \times S^1$.
Studying the winding sector of the twisted $\mathbb R^d \times S^1$  worldsheet theory, we were able to derive \eqref{eq:hag_flat_final}. 
As we explain, it is also possible to derive \eqref{eq:hag_flat_final} perturbatively in $\nu$ from a target space analysis. Using a Kaluza-Klein (KK) reduction, we find a novel ($\nu$ dependent) effective field theory (EFT) for the thermal winding mode on $
\mathbb{R}^d$. In this way, the Hagedorn temperature \eqref{eq:hag_flat_final} is the temperature at which a winding mode turns massless.

In section \ref{sec:sss}, we follow~\cite{Horowitz:1997jc} and employ this $d$-dimensional EFT of the winding mode coupled to gravity to find a normalizable winding condensate, or a rotating string star.\footnote{Recently,~\cite{Santos:2024ycg} presented an attempt to find a rotating winding condensate. To our understanding, their underlying assumption is significantly different from ours. Most importantly, the asymptotic geometry of their solution was $\mathbb{R}^d \times S^1_\beta$ with no asymptotic angular velocity ($\nu=0$). As a result, the authors also introduce an intrinsic spin to the winding profile, and work only with heterotic string theory (in which the winding mode current sources the KK vector). Due to the asymptotic angular velocity, our equations of motion differ at leading order. Our winding profile has no intrinsic spin and is invariant to the remaining $SO(2)\times SO(d-2)$ rotation symmetry. While we consider, for simplicity, only bosonic and type II string star configurations, we believe a solution for heterotic strings exists and will share similar properties.}
To find the solution, we employ perturbation theory in $\nu \ll 1$, which to leading order amounts to a rigid-body approximation.
By analytically continuing the solution back to real angular velocity ($2\pi \nu i = \beta \Omega$), we write down the thermodynamic properties of the rotating string star saddle.
The rotating string star shares many properties with the non-rotating solution, such as $O(G_N^{-1})$ entropy. The winding profile is invariant to the remaining $SO(2)\times SO(d-2)$ rotation symmetry group and thus has no intrinsic spin.
Notably, this solution still carries a non-zero $O(G_N^{-1})$ angular momentum, due to the non-trivial $\nu$ dependence, similar to the rotating black hole.
In section \ref{sec:corr}, we compare these properties to rotating black holes and to free strings. As in the non-rotating case, we find qualitative agreement between both in their intermediate parameter regimes.

Finally, in section \ref{sec:ads} we follow~\cite{Urbach:2022xzw,Urbach:2023npi} and discuss the extension of our results to anti-de Sitter (AdS) space. We compute the leading behavior of the AdS Hagedorn temperature $R_H(\nu)$ in $l_s/l_{ads}$, and comment on the expected properties of the rotating AdS string star.

\section{The thermodynamics of highly-excited rotating strings}
\label{sec:flat_space_therm}
We begin by discussing the thermodynamic properties of free rotating strings at high energies. We will consider either bosonic, type II, or heterotic (critical) string theories in flat space, all at zero string coupling $g_s=0$. In this limit, the canonical partition function is well-defined~\cite{Atick:1988si} and the free string phase is dominant for arbitrarily high energies.
We denote the number of spatial directions by $d$ ($d=25$ in the bosonic case, and $d=9$ for the superstring) and the plane of rotation by coordinates $X^1,X^2$. Denoting the worldsheet spatial coordinate by $\sigma \sim \sigma+l$, the worldsheet angular momentum generator is given by (for the superstring, there is a second fermionic term)
\begin{equation}
    J = \frac{1}{2\pi \alpha'}\int_0^l d\sigma \left(X^1 \partial_\tau X^2 - X^2 \partial_\tau X^1\right).
\end{equation}
In this normalization, the angular momentum is dimensionless and integer-quantized for bosonic excitations.

\subsection{The microcanonical ensemble}\label{subsec:micro}
Fixing the mass $M$ and the angular momentum $J$, we would like to count the number of microstates $S(M,J)$ for high energies $\alpha' M^2 \gg 1$. For a recent review about free rotating strings in the microcanonical ensemble, see~\cite{Ceplak:2023afb,Ceplak:2024dxm}. In the high-energy limit, the string theory partition function is dominated by the single string partition function~\cite{Mertens:2015ola}. Therefore, without loss of generality, we will focus on the single-string entropy $S(M,J)$.
For a given mass $M$, the maximal possible amount of angular momentum is constructed by appropriately exciting the $X^1$, $X^2$ oscillators. The mass shell in this extreme case is given by $\frac{1}{2} \alpha' M^2 = J + c$, where $c$ is a theory-dependent fixed number. Therefore, at high energies $\alpha' M^2 \gg 1$, the angular momentum is bounded by the Regge trajectory
\begin{equation}\label{eq:J_bound}
    J \le J_\text{Regge} = \frac{1}{2} \alpha' M^2 + O(M^0).
\end{equation}

Considering the high-energy entropy, one needs to specify the scaling of the angular momentum in the high-energy limit. One possible scaling is to simply fix $J$ as we take $\alpha' M^2 \gg 1$. In this limit, the leading order entropy has linear growth, 
\begin{equation}\label{eq:hag}
    S(M,J) = \beta_H M + O(\log M, M^0),
\end{equation}
as in the standard single-string entropy $S(M)$~\cite{Huang:1970iq,Gross:1985fr,Bowick:1985az,Atick:1988si}.
The linear growth of the entropy signals a limiting temperature for the system $T_H = 1/\beta_H$, termed the Hagedorn temperature~\cite{Hagedorn:1965st}. It is useful to write down the temperature in terms of the ``thermal radius'' $R_H = \beta_H/(2\pi)$,
\begin{equation}\label{eq:hag_nu_zero}
    R_H / l_s = 
    \begin{cases}
        2, & \text{Bosonic}\\
        \sqrt{2}, & \text{Type II}\\
        \frac{1}{2}\left(2+\sqrt{2}\right),& \text{Heterotic}
    \end{cases}.
\end{equation}

Based on the aforementioned bound \eqref{eq:J_bound}, a more natural scaling is to keep $-1/2<J/(\alpha' M^2)<1/2$ fixed as we take the high-energy limit.
As we show in appendix \ref{app:micro_entropy} (see also~\cite{Matsuo:2009sx}), the high energy behavior is 
\begin{equation}\label{eq:S_M_J}
    S(M,J)= 2\pi \tilde R_H \sqrt{M^2-2J/\alpha'} + O(M^0).
\end{equation}
This result has a very appealing interpretation. An $\sqrt{2J}/l_s$ amount of the energy is spent to create a state with angular momentum $J$, while the rest of the energy $\sqrt{M^2-2J/\alpha'}$ behaves as in \eqref{eq:hag}. In other words, this scaling also has Hagedorn growth of the entropy, with a Hagedorn radius (see appendix \ref{app:micro_entropy})
\begin{equation}
\begin{split}
    \tilde R_{H}/l_s =
    \begin{cases}
        2, & \text{Bosonic}\\
        \sqrt{2}, & \text{Type II} \\
        \sqrt{3}, & \text{Heterotic}.
    \end{cases}.
\end{split}
\end{equation}
Notice that the Hagedorn radius remains the same for bosonic and type II theories, but is altered for heterotic strings. The reason is that for large $J$, the heterotic oscillators' occupation numbers are distributed differently between the worldsheet right and left movers to maximize the entropy, leading to a different exponential growth of states.

A third possible high-energy ($l_s M \gg 1$) limit is to keep $J/(l_s M)$ fixed. We can further expand in both large $l_s M \gg 1$ and small $J/(l_s M) \ll 1$. For a derivation of the analysis, see appendix \ref{app:J/M free string} and~\cite{Russo:1994ev} for more details. 
The entropy has the double expansion 
\begin{equation}\label{eq:entropy_J_M}
    S(M,J) = \beta_H M - \alpha_1 \log(l_s M) + \mathcal{O}(M^0) - \alpha_2 \frac{J^2}{\alpha' M^2}\left(1+\mathcal{O}(M^{-1})\right) + \mathcal{O}\left(\frac{J^4}{(\alpha')^2 M^4}\right),
\end{equation}
with $\alpha_1,\alpha_2$ theory-dependent constants\footnote{In case where some of the dimensions are compact, this formula needs to be corrected~\cite{Mertens:2015ola}.}
\begin{equation}\label{eq:a1a2}
    \alpha_1 = \begin{cases}
        27, & \text{Bosonic}\\
        15, & \text{Type II}\\
        21,& \text{Heterotic}
    \end{cases}, \qquad 
    \alpha_2 = 2(2\pi)^2
    \begin{cases}
        \frac{1}{2}, & \text{Bosonic}\\
        \frac{1}{4}, & \text{Type II}\\
        \frac{1}{3},& \text{Heterotic}
    \end{cases}.
\end{equation}
As we can see, the leading behavior is still Hagedorn, while the angular momentum contributed negatively, following the same intuition we saw in \eqref{eq:S_M_J}.

Following~\cite{Ceplak:2024dxm}, we would also like to estimate the shape of the rotating string. As an estimate for the typical sizes of the string in the rotation plane ($L_{||}$) and in the directions orthogonal to it ($L_\perp$), one can calculate the variance of the corresponding coordinate $\langle (X^i)^2 \rangle$. 
In appendix \ref{app:J/M free string}, we analyze the case of bosonic strings, which gives
\begin{equation}\label{eq:L_J_M}
    \begin{split}
        L_{||} & = \frac{l_s^{\frac{3}{2}} \sqrt{\pi M}}{\sqrt{12}}\left(1+\mathcal{O}(M^{-1})+\frac{12J^2}{5(l_s M )^2}+\mathcal{O}\left(\frac{J^4}{M^4}\right) \right) \\
        L_{\perp} & = \frac{l_s^{\frac{3}{2}} \sqrt{\pi M}}{\sqrt{12}}\left(1+\mathcal{O}(M^{-1})  + \mathcal{O}\left(\frac{J^2}{M^3}\right) +\mathcal{O}\left(\frac{J^4}{M^4}\right)\right).
    \end{split}
\end{equation}
At leading zeroth order in $J/(l_s M)$, the sizes agree and coincide with the random walk approximation $L^2/\alpha' \sim l_s M$. At subleading order, we find $L_{||} > L_\perp$, following the Newtonian expectation due to the centrifugal force.
It will be instructive to define the eccentricity $e = \sqrt{1-\frac{L_{\perp}^2}{L_{||}^2}}$ of the shape, leading to
\begin{equation}\label{eq:ecc_J_M}
    e = 2\sqrt{\frac{6}{5}}\frac{J}{l_s M}\left(1 + \mathcal{O}(M^{-1})+ \mathcal{O}\left(\frac{J^2}{M^2}\right)\right).
\end{equation}
Finally, we can compute the angular velocity $\Omega$ (the variable canonically conjugate to $J$) to leading order in the angular momentum $J$
\begin{equation}\label{eq:omega_J_M}
    l_s \Omega = \frac{6}{\pi} \frac{J}{\alpha' M^2}\left(1 + \mathcal{O}(M^{-1})+ \mathcal{O}\left(\frac{J^2}{\alpha' M^2}\right)\right). 
\end{equation}
The leading order scaling can be understood in terms of the rigid-body approximation: for the non-rotating string, the random-walk approximation gives the typical size $\sqrt{M}$. As a result, the moment of inertia is $\mathcal{I} \sim M^2$. Using the rigid body result $\Omega \sim J/ \mathcal{I} \sim J/(\alpha' M^2)$, we find the same scaling as \eqref{eq:omega_J_M}. 
Although the calculation of the shape was done only for bosonic strings, we believe they are qualitatively the same also for type II and heterotic strings.

\subsection{The grand canonical ensemble}\label{subsec:grand_can}
We would like to add a chemical potential for the angular momentum. One way to do it is to add a potential term in the microcanonical ensemble, of the type
\begin{equation}
    Z(M,\mu) = \sum_{|J|\leq \frac{1}{2} \alpha'M^2} e^{S(M,J)} e^{-\mu J}.
\end{equation}
However, the high-energy behavior of this ensemble is quite mundane. In the high energy limit $\alpha' M^2 \gg 1$ (at fixed $\mu$), a maximal $S(M,J)-\mu J$ is achieved by maximizing the $-\mu J$ term, which dominates over the entropy term. 
Therefore, the ensemble is dominated at high energies by maximally rotating (or extremal) strings \eqref{eq:J_bound}.
A more natural ensemble is the grand canonical ensemble with fixed (inverse) temperature $\beta$ and angular velocity $\Omega$:
\begin{equation}\label{eq:Z_beta_Omega}
    Z(\beta,\Omega) = \text{Tr} \left( e^{-\beta (H-\Omega J)} \right).
\end{equation}
This ensemble is especially familiar in the context of AdS/CFT~\cite{Hawking:1998kw,Hawking:1999dp}.
Unfortunately, in flat space, fixing the angular velocity without fixing the energy is generically unstable. The reason is that it is always possible to excite the system with low energy but arbitrarily high angular momentum $\Delta J$, so that the Boltzmann factor will be enhanced by $\sim \exp(+\Omega \ \Delta J)$ \cite{Witten:2021nzp,Kim:2023sig}. Allowing arbitrarily high $\Delta J$ will result in a diverging partition function.%
\footnote{In other words, the multi-string partition function is divergent. As a matter of fact, the single-string partition function is also divergent for a somewhat different reason. As we explained above, for high enough energies, the system is dominated by maximally rotating strings. While the Boltzmann factor is $\exp(-\beta M)$, the potential enhances it by  $\sim \exp(\beta \Omega M^2/2)$. Therefore, the sum over $M$ results in a diverging partition function. See appendix \ref{app:newton} for a classical intuition for this result.}

One possible approach to cure this exponential divergence is to set an imaginary angular velocity. 
Substituting $\beta \Omega = 2\pi i \nu$ in \eqref{eq:Z_beta_Omega} leads to the partition function
\begin{equation}\label{eq:grand_nu}
    Z(\beta,\nu)= \text{Tr} \left( e^{-\beta H  +2\pi i \nu J} \right).
\end{equation}
While it might look like an odd thing to do, note that, for example, in AdS$_3$~\cite{Maldacena:1998bw,Maloney:2007ud} this construction is natural, and dual to the familiar CFT$_2$ torus partition function. Similarly, it is in this ensemble that the rotating black hole solution is real (in Euclidean signature).
On the other hand, this ensemble is harder to interpret statistically, as it gives complex probabilities for microstates.

The first thing to note about this ensemble is that the imaginary angular velocity $\nu$ is actually compact.
Since for the bosonic theory states have integer $J$, the partition function is periodic in $\nu$, namely $Z(\beta,\nu) = Z(\beta,\nu+1)$. Due to half-integer $J$, for superstring theories we expect $Z(\beta,\nu) = Z(\beta,\nu+2)$.
Secondly, an imaginary angular velocity also means that the canonical ensemble is no longer dominated by maximally rotating strings. This is because instead of exponential growth, the Boltzmann factor only acquires a phase $\exp(2\pi i \nu J)$.
This opens the possibility that at fixed $\nu$ the high energy behavior will be dominated by Hagedorn growth of the entropy, accompanied by the phases coming from the angular momentum term.
The main result of this section is that this ensemble indeed possesses a non-trivial Hagedorn instability. More precisely, we claim that the partition function $Z(\beta,\nu)$ diverges above some temperature $\beta < \beta_H(\nu)$, for some function $\beta_H(\nu)$. 

In the remainder of this section, we will show that for bosonic strings. We refer the reader to appendix \ref{app:state_count} for a derivation also for type II and heterotic strings.
Our reasoning will follow an analysis of~\cite{Russo:1994ev,Matsuo:2009sx}, or more recently by~\cite{Ceplak:2024dxm}. 
We begin by working in light-cone quantization, and computing the worldsheet partition function with fixed worldsheet temperature $1/b$ and imaginary (target space) angular velocity $\nu$. The (single chirality) result for the $d$-dimensional bosonic string is ($x=e^{-b}$)\cite{Matsuo:2009sx}
\begin{equation}
\begin{split}
    Z(b,\nu) &= \prod_{n=1}^\infty \left(\frac{1}{1-x^n}\right)^{d-3} \frac{1}{(1-e^{2\pi i \nu} x^n)(1-e^{-2\pi i \nu} x^n)}\\
    &= 
    \frac{-2 e^{-\frac{d-1}{24}b} \sin(\pi\nu)}{\eta^{d-4}(i b/(2\pi)) \vartheta_{11}(\nu,i b/(2\pi))}.
\end{split}
\end{equation}
With $\eta$ the Dedekind function and $\vartheta_{11}$ the Jacobi $\theta$ function~\cite{Polchinski:1998rq}. Using the modular properties of the expression, the small $b$ behavior is
\begin{equation}
    Z(b,\nu) = \exp(-\frac{2\pi^2}{b} a(\nu) + O(\log b)),
\end{equation}
with $a(\nu) = 4+2\nu(\nu-1)$ at $d=25$ and the fundamental region $0\le \nu <1$. 
This behavior of $Z(b,\nu)$ signals that at fixed worldsheet level $N$, the partition function behaves as 
\begin{equation}
        Z(N,\nu) = \int \frac{db}{2\pi i} e^{N b} Z\left(b,\nu\right) \sim \exp\left(\sqrt{8\pi^2 a(\nu) N}+O(N^0)\right).
\end{equation}
To reach the RHS, we used a saddle point approximation, which gives the leading piece in the large $N$ (small $b$) limit. Considering the two chiralities and the on-shell relation $\alpha' M^2 = N/4 + O(1)$, we find the Hagedorn growth
\begin{equation}
    Z(M,\nu) \sim \exp (\sqrt{4\pi^2 a(\nu) \alpha'} M).
\end{equation}
Defining the Hagedorn radius $R_H = \beta_H/(2\pi)$ gives
\begin{equation}\label{eq:R_H_B}
    R_H^2(\nu)/\alpha' = a(\nu) = 4 + 2 \nu \left(\nu-1 \right).
\end{equation}
The linear growth of $\log Z(M,\nu)$ signals that the target-space canonical partition function $Z(\beta,\nu)$ \eqref{eq:grand_nu} will indeed diverge for $\beta < \beta_H(\nu)$.
At $\nu=0$, this result coincides with the standard Hagedorn temperature \eqref{eq:hag_nu_zero}. As we increase $\nu$, the Hagedorn temperature initially increases, and then, as required, goes back to itself at $\nu=1$. The initial increase of the Hagedorn temperature is because the $\nu$-dependent phases introduce some cancellations that suppress the divergence.

\subsection{A Euclidean derivation of \titlemath{$\beta_H(\nu)$}}
\label{subsec:euc_hag}

Following the work of~\cite{Sathiapalan:1986db,Kogan:1987jd,Atick:1988si}, there is a second, intuitive way to explain the Hagedorn instability of string theory in the canonical ensemble. The idea is to follow~\cite{Polchinski:1985zf}, and understand the canonical partition function $Z(\beta)$ as the Euclidean string theory partition function on $\mathbb{R}^d \times S^1_\beta$ (with anti-periodic boundary condition for the fermions).
In terms of the Euclidean theory, the Hagedorn instability is a divergence of the 1-loop path integral in this background. Unlike field theory, string theory in this background includes sectors of strings winding around the thermal circle. Due to their extension along the thermal circle, the mass of the winding strings is, up to an overall constant, proportional to $m^2(\beta) \sim \beta^2$. A direct computation shows that the lightest winding mode turns massless at the  Hagedorn temperature $\beta_H$, and tachyonic for higher temperatures. As it becomes massless, the 1-loop partition function acquires a new divergence at $\beta=\beta_H$, and turns ill-defined for $\beta < \beta_H$. This is the Euclidean reasoning behind the linear Hagedorn growth of the entropy.

In this section, we would like to extend this discussion to the grand canonical ensemble with non-zero imaginary angular velocity $\nu$. As a first step, we would like to cast the grand canonical partition function \eqref{eq:grand_nu} in terms of a Euclidean partition function. The metric on $\mathbb{R}^d\times S^1_\beta$ can be written as
\begin{equation}\label{eq:WS_metric}
\begin{split}
    ds^2 &= R^2 (dX^0)^2 + \sum_{i=1}^d (dX^i)^2 \\
    &= R^2 (dX^0)^2 + |d\Phi|^2 + \sum_{i=3}^d (dX^i)^2,
\end{split}
\end{equation}
with $R=\beta/(2\pi)$, and $X^1,X^2$ spanning the rotation plane of $J$. In the second line, we have introduced the complex scalar $\Phi = X^1 + i X^2$. 
Due to the trace in \eqref{eq:grand_nu}, we identify
\begin{equation}\label{eq:WS_ident}
    (X^0,\Phi) \sim (X^0 + 2\pi, e^{-2\pi i \nu} \Phi).
\end{equation}
Following~\cite{Polchinski:1985zf}, the trace \eqref{eq:grand_nu} is simply equal to the string theory partition function on the background \eqref{eq:WS_metric} under the identification \eqref{eq:WS_ident}.

Treating $X^0$ as a spatial circle, these spaces are known as Melvin spaces. While originally proposed as solutions of Einstein-Maxwell theory~\cite{Melvin:1963qx}, they were later embedded in KK theory~\cite{Gibbons:1987ps,Dowker:1993bt}. The KK reduction of \eqref{eq:WS_metric} includes a non-trivial (but asymptotically flat) curved metric, dilaton, and KK magnetic flux profile localized around the center of the rotation plane $\Phi=0$ (also known as a ``magnetic fluxbrane''). As an exact Lorentzian string theory model, it was discussed, for example, in ~\cite{Tseytlin:1994ei,Dowker:1995sg,Russo:1995tj,Gutperle:2001mb}.\footnote{We thank Roberto Emparan for introducing us to this subject.} To the best of our knowledge, the following thermal interpretation we suggest is novel. However, it is interesting to note that the possible relation between this background and a rotating black hole was already discussed in~\cite{Dowker:1995gb}.

We will begin by discussing bosonic string theory ($d=25$).
In the coordinates \eqref{eq:WS_metric}, the worldsheet theory is simply a free theory for the $X$s, orbifolded by the identification \eqref{eq:WS_ident}.
This theory is a generalization of the thermal theory ($\nu=0$) to any other $\nu$.
Here as well, the orbifold theory includes twisted sectors of strings winding around the thermal circle.
In the first winding sector, the boundary condition for the fields is 
\begin{equation}
    X^0(\sigma + l) = X^0(\sigma) + 2\pi, \quad \Phi(\sigma + l) = e^{2 \pi i\nu} \Phi(\sigma).
\end{equation}

We use light-cone quantization, choosing two of the other $d-2$ directions for gauge-fixing~\cite{Polchinski:1998rq}. To find the spectrum of the winding sector, we need to determine the worldsheet ground-state energy in that sector.
The contribution of $X^0$ and all the other $d-2$ coordinates to the ground-state energy is standard.  The complex scalar $\Phi$, however, has twisted boundary conditions which shift the standard harmonics from  $\omega_n=|n|$ to $\omega_n = |n+\nu|$.
As a result, the $\Phi$ contribution to the ground-state energy is\footnote{Using the identity~\cite{Polchinski:1998rq}
\begin{equation}
    \sum_{n=1}^\infty (n+\nu) = -\frac{1}{12} -\frac{1}{2}(\nu+\nu^2).
\end{equation}
}
\begin{equation}\label{eq:casimir_B}
    E_{0,\Phi} = \sum_{n=-\infty}^{\infty} |n + \nu| = - \frac{1}{6} - \nu \left(\nu -1\right),
\end{equation}
where in the second equality we assumed the fundamental region $0\le\nu<1$.
Together with the standard contributions to the ground-state energy, we find the mass-shell condition
\begin{equation}\label{eq:mass_shell_B}
    \frac{1}{4}\alpha' M^2 
    = \frac{R^2}{4\alpha'} -1- \frac{1}{2}\nu\left(\nu-1\right) + N,
\end{equation}
with $N$ the level.
As we explained above, the Hagedorn temperature is the temperature at which the lightest winding mode in \eqref{eq:mass_shell_B} turns massless. Since the lightest physical mode has $N=0$, setting $M^2=0$ leads to
\begin{equation}
    R^2_{H}/\alpha' = 4 + 2\nu(\nu-1).
\end{equation}
This is exactly what we found in \eqref{eq:R_H_B} using state counting!

For superstring theories, we also need to twist the worldsheet fermions $\psi_1$ and $\psi_2$. Following the bosonic discussion, we denote the complex spinor $\Psi = \psi^1+i \psi^2$. As we are interested in the (twisted) NS sector, the boundary condition for the complex spinor is
\begin{equation}
    \Psi(\sigma + l) = e^{-2\pi i(\nu+1/2)} \Psi(\sigma).
\end{equation}
Summing over the harmonic modes, the ground-state energy for $\Psi$ is given by
\begin{equation}\label{eq:casimir_F}
    E_{0,\Psi} = -\sum_{n=-\infty}^\infty \left|n+\nu+\frac{1}{2}\right| = -\frac{1}{12}+\nu^2,
\end{equation}
where this time we assumed the region $-1/2<\nu<1/2$. Notice that this expression has periodicity $\nu \sim \nu+1$.

For type II string theory, we use \eqref{eq:casimir_B},\eqref{eq:casimir_F} to get the mass-shell condition (for $-1/2<\nu<1/2$)
\begin{equation}\label{eq:mass_shell_ii}
    \frac{1}{4}\alpha' M^2  
    =\frac{R^2}{4\alpha^{'}} - \frac{1}{2} + \frac{1}{2}|\nu| + N.
\end{equation}
To find the mass of the lightest winding mode, we need to ask what the lowest allowed level $N$ is, as a function of $\nu$.
At $\nu=0$,~\cite{Atick:1988si} showed that we should take $N=0$. At $\nu=1$, however, the target-space fermions are periodic on the thermal circle. Using T-duality, the small $R$ limit is dual to the decompactified $\mathbb{R}^{10}$ limit. In this limit, we know that the tachyon $N=0$ is projected out by GSO, and the lightest state has $N=1/2$~\cite{Polchinski:1998rr}. What is the correct value for $N$?
Another way to phrase the question is the following. Although the full theory is expected to go back to itself only under $\nu\sim\nu+2$, the mass-shell condition \eqref{eq:mass_shell_ii} appears to have a shorter periodicity of $\nu \sim \nu+1$, as both  \eqref{eq:casimir_B} and \eqref{eq:casimir_F} have that periodicity. How is that possible?

The answer is that the GSO projection and the $\nu\mapsto \nu+1$ transformation are not commutative. Notice that at $\nu=\pm 1/2$ the $n=\mp 1$ creation and annihilation operators switch roles. For this reason, for $1/2<|\nu|\le 1$ the ground-state's fermion number changes sign and the lightest physical state is an excited state. Taking this into account gives the $\nu$-dependent minimal level
\begin{equation}\label{eq:level_ii}
    N = \begin{cases}
        0,& 0<|\nu|<1/2\\
        |\nu|-\frac{1}{2},& 1/2<|\nu|<1
    \end{cases}.
\end{equation}
Therefore, the lightest winding mode for type II has a mass
\begin{equation}\label{eq:mass_ii}
    \alpha' M^2 
    =\frac{R^2}{\alpha^{'}} - 2 + 2|\nu|.
\end{equation}
This time, due to \eqref{eq:level_ii}, \eqref{eq:mass_ii} holds for the full fundamental region $-1\le \nu\le 1$, with periodicity $\nu \sim \nu+2$. 

Setting the mass to zero gives the type II Hagedorn temperature ($-1\le \nu\le 1$)
\begin{equation}\label{eq:ehag_II}
    R^2_{H}/\alpha' = 2 (1- |\nu|).
\end{equation}
As we show in appendix \ref{app:state_count}, one can obtain the same Hagedorn growth from direct state counting.
At $\nu=0$, we get back the standard answer $R_H=\sqrt{2}  l_s$ \eqref{eq:hag_nu_zero}~\cite{Atick:1988si}. As in the bosonic case, we find that increasing $\nu$ initially increases the Hagedorn temperature to leading order. This time, however, we find that at $\nu=1$ the Hagedorn temperature is infinity ($R_H=0$)!  
We can understand this result as follows.
At $\nu=1$, the target space has periodic boundary conditions for the fermions. In the $R\ll l_s$ limit, this mode is T-dual to a graviton momentum mode of $g_{00}$. We expect such a mode to turn massless $m^2=0$ in the non-compact $R=0$ limit, which gives $R_H=0$ in this case.
In fact, due to the periodic boundary conditions, at $\nu=1$ we have $e^{2\pi i \nu J}=(-1)^F$ (where $F$ is the target space fermion number), and the partition function actually computes the supersymmetric index of the theory~\cite{Urbach:2025eyu}.
Due to cancellations inside the multiplets, we don't expect the computation to depend on the temperature, and specifically, there is no Hagedorn instability. As a result, \eqref{eq:ehag_II} should strictly hold only for $\nu \ne 1 \text{ mod }2$.

Finally, we turn to heterotic strings. At $\nu=0$, the GSO projection dictates that the momentum along $X_0$ is an odd half-integer (instead of an integer)~\cite{Atick:1988si}. To generalize the analysis to $\nu \ne 0$, we keep the momentum $n$ unfixed and write down the right and left mass shell conditions. For the right-movers, we have the twisted boson \eqref{eq:mass_shell_B} together with $32$ untwisted fermions
\begin{equation}
    \frac{1}{4}\alpha' M^2 = -1 + \frac{1}{2}\left(|\nu|-\nu^2\right) + N + \frac{1}{4\alpha'}\left(R + \frac{n \alpha'}{R}\right)^2.
\end{equation}
For the left-movers, we have the same condition as in type II \eqref{eq:mass_ii}
\begin{equation}
    \frac{1}{4}\alpha' M^2 = - \frac{1}{2} + \frac{1}{2}|\nu| + \tilde N + \frac{1}{4\alpha'}\left(R - \frac{ n \alpha'}{R}\right)^2.
\end{equation}
In both cases, we assumed $-1/2<\nu<1/2$ with $\nu \sim \nu+1$.
Taking the difference, we find
\begin{equation}
    n = \tilde N- N + \frac{1}{2} +\frac{1}{2}\nu^2.
\end{equation}
Note that due to the shifted frequencies of $\Phi$ and $\Psi$, $\tilde N-N$ is generally not an integer.

What are the lightest physical levels $N,\tilde N$? At least for $R \gg l_s$, we can ignore the momentum dependence in the mass shell condition, and ask for the minimal $N,\tilde N$ separately. Following the discussion above, the lowest level for bosonic strings is $N=0$, and \eqref{eq:level_ii} for $\tilde N$. From the mass shell relation, we find that this mode has momentum and mass ($-1<\nu<1$) 
\begin{equation}\label{eq:mass_shell_het}
    n=\frac{1+\nu^2}{2}, \quad
    \alpha'M^2=
    \frac{R^2}{\alpha'}+
    \frac{\alpha'}{R^2}\frac{(1+\nu^2)^2}{4}
    -\nu^2+2|\nu|-3.
\end{equation}
As can be checked explicitly, this mode remains the lightest winding excitation also for smaller radii $R$, until the mode turns massless. Solving \eqref{eq:mass_shell_het} with $M^2=0$ gives the heterotic Hagedorn temperature 
\begin{equation}\label{eq:ehag_het}
    R_H/l_s = \frac{1}{2}\left(\sqrt{4-2|\nu|+2\nu^2}+\sqrt{2-2|\nu|}\right).
\end{equation}
For a direct state-counting derivation of this Hagedorn temperature, see appendix \ref{app:state_count}.
As a sanity check, note that this result coincides with the standard $R_{H}/l_s = 1 + 1/\sqrt{2}$ at $\nu=0$~\cite{Gross:1985fr}. As in the type II case, there is no Hagedorn divergence or instability at $\nu=1$ due to supersymmetric cancellations, and \eqref{eq:ehag_het} should be strictly understood only for $\nu \ne 1 \text{ mod } 2$.\footnote{
Note that here, unlike type II, we have a finite ``Hagedorn temperature'' $R_H=l_s$ at $\nu=1$.
The behavior of the winding mode and the string star solution close to $\nu=1$ will be discussed elsewhere~\cite{Urbach:2025eyu}.}

\subsection{A target space analysis}\label{subsec:Atick_W}
We would like to explain the results of the previous section also from a target space perspective. At $\nu=0$, the target space perspective was put forward by Atick and Witten~\cite{Atick:1988si}. Reducing on the thermal $S^1_R$ for $R-R_H \ll l_s$, the Euclidean string theory includes a light particle associated with the (lightest) winding mode, with the $d$-dimensional quadratic action
\begin{equation}\label{eq:eft_chi2}
    S = \int d^d x \left(|\partial \chi|^2 + m^2 |\chi|^2 \right).
\end{equation}
Following the string theory analysis, the mass is an $R$-dependent function. It is instructive to write the mass in terms of the metric components
\begin{equation}\label{eq:chi_mass}
    m^2 = \frac{G_{tt}}{(\alpha')^2}+n^2 G^{tt}-\frac{c}{\alpha'},
\end{equation}
where we denote $t \sim t+2\pi$ to be the direction in which the winding mode winds. In terms of \eqref{eq:WS_metric}, we have $\partial_t = \partial_0$, which leads to 
\begin{equation}
    G_{tt}=R^2, \quad G^{tt}=R^{-2}.
\end{equation}
The mass parameters $c,n$ (which correspond to the worldsheet Casimir energy and the temporal momentum, respectively) depend on the string theory in question, and are given by
\begin{equation}\label{eq:wind_param}
    n = \begin{cases}
        0, & \text{Bosonic}\\
        0, & \text{Type II}\\
        \frac{1}{2},& \text{Heterotic}
    \end{cases}, \quad
    c = \begin{cases}
        4, & \text{Bosonic}\\
        2, & \text{Type II}\\
        3,& \text{Heterotic}
    \end{cases}.
\end{equation}
The Hagedorn instability takes place when the 1-loop partition function diverges. In this language, it happens when $\chi$ becomes massless.

What would be the effective theory once we add an imaginary angular velocity $\nu\ne 0$? To understand what is going on, it will be useful to write down the background metric in the following way. 
Starting from \eqref{eq:WS_metric}, we move to cylindrical coordinates
\begin{equation}
    ds^2 = R^2 (dX^0)^2 + \rho^2 d\varphi^2 + d\rho^2 +  \sum_{i=3}^d \left(dX^i\right)^2,
\end{equation}
with $\Phi = \rho e^{i \varphi}$. In those coordinates, we identify $(X^0,\varphi) \sim (X^0+2\pi, \varphi-2\pi \nu) \sim (X^0,\varphi+2\pi)$.
Now define $t=X^0$ and $\phi = \varphi+\nu t$. In these coordinates, $t\sim t+2\pi$ and $\phi \sim \phi+2\pi$ with no coupling in their identification. The metric in these coordinates takes the form
\begin{equation}\label{eq:new_flat_metric_1}
\begin{split}
    ds^2 &= R^2 dt^2 + \rho^2 (d\phi - \nu dt)^2 + d\rho^2 +  \sum_{i=3} \left(dX^i\right)^2\\
    &= (R^2 + \nu^2 \rho^2) \left(dt - A_\phi d\phi\right)^2 + \frac{\rho^2}{1+\frac{\nu^2 \rho^2}{R^2}}d\phi^2 + d\rho^2+ \sum_{i=3}^d \left(dX^i\right)^2.
\end{split}
\end{equation}
In the second line, we wrote the metric in a Kaluza-Klein (KK) form, with the KK vector $A_\phi = \frac{\nu \rho^2}{R^2+\nu^2 \rho^2}$. As mentioned in the previous section, this is a Euclidean version of the Melvin magnetic fluxbrane space. 

Performing a KK reduction to this metric will give a local EFT for modes with zero momentum on the time circle. To find the new Hagedorn temperature, we need to write down the new quadratic theory for the winding mode $\chi$ on this background.
Up to order $O(\alpha'^{0})$, the only correction to \eqref{eq:eft_chi2} is to turn the derivatives into covariant derivatives with respect to $A_i$, with charge $n$ (the thermal momentum). Assuming azimutal symmetry for $\chi$, this term gives back \eqref{eq:chi_mass} as the effective mass, only with the new metric \eqref{eq:new_flat_metric_1},
\begin{equation}
    G_{tt} = R^2 + \nu^2 \rho^2, \quad G^{tt} = R^{-2}.
\end{equation}

We found that up to order $O(\alpha'^0)$, the only new term in the quadratic action for $\chi$ is a universal (theory-independent) mass term $\frac{\nu^2}{(\alpha')^2} \rho^2 |\chi|^2$. Turning the angular velocity to be real again, we find a negative $-\Omega^2 \frac{R^2}{(\alpha')^2} \rho^2 |\chi|^2$ mass term. This term destabilizes the action at large radii $\rho$. This is precisely the Euclidean avatar of the fact that the fixed $\Omega$ ensemble is divergent (see section \ref{subsec:grand_can}). As we show in appendix \ref{app:newton}, the same sort of mass term appears for this ensemble also for classical particles.

Since $\nu$ appears in the effective action through the metric components, the $\alpha'$ expansion corresponds to the Taylor expansion of the physical mass in orders of $\nu$. To show an exact match between the worldsheet result and the target space, one would need to properly find all the higher $\alpha'$ couplings of the KK vector to the winding mode (for each string theory separately). One such coupling is the term $\alpha' F^2 |\chi|^2$, with $F$ the field strength of the KK vector. Since we only consider the quadratic theory to order $O(\alpha'^0)$, we can only use it to find the $O(\nu)$ correction to the physical mass.
In all the string theories we considered in the previous section \eqref{eq:mass_shell_B}, \eqref{eq:mass_shell_ii}, \eqref{eq:mass_shell_het}, this leading correction to the physical mass had the form
\begin{equation}\label{eq:mass_corr}
    M^2(\nu) = M^2_0 + \frac{2|\nu|}{\alpha'} + O(\nu^2).
\end{equation}
In the rest of the section, we will show how the $O(\alpha^{\prime 0})$ EFT exactly reproduces \eqref{eq:mass_corr}.

At quadratic order \eqref{eq:eft_chi2}, the zero mode solution for $\chi$ satisfies the equation $\nabla^2 \chi - m^2 \chi=0$. Assuming $\chi$ depends only on $\rho$\footnote{As we are looking for the lightest normalizable solution, it will be rotationally symmetric. As for the other $d-2$ non-compact directions $\vec X$, the zero mode is simply a constant, just like in the standard flat space computation $\nu=0$.}, it can also be written as a Schrodinger equation
\begin{equation}\label{eq:sch_eq}
\begin{split}
    -\chi''(\rho) - \frac{1}{\rho} \chi'(\rho)+\frac{\nu^2}{(\alpha')^2} \rho^2 \chi(\rho) = E \chi(\rho),
\end{split}
\end{equation}
with ``energy eigenvalue'' $E$
\begin{equation}\label{eq:my_E}
    E = \frac{c }{\alpha'}-\frac{R^2}{(\alpha')^2}-\frac{n^2}{R^2}.
\end{equation}
The lightest normalizable eigenfunction of \eqref{eq:sch_eq} is $\chi(\rho) = \exp(-|\nu| \rho^2/(2\alpha'))$. This solution has energy $E=2|\nu|/\alpha'$, exactly matching the worldsheet result \eqref{eq:mass_corr} of the previous section. Correctly reproducing the $O(\nu)$ correction to the mass, this result also reproduces the right $O(\nu)$ correction to the Hagedorn temperature \eqref{eq:hag_flat_final} by solving $E=2|\nu|/\alpha'$ for $R$:
\begin{equation}
\begin{split}
    R_H/l_s &= \frac{1}{2}\left(\sqrt{c+2n}+\sqrt{c-2n}\right) - \frac{1}{2}\left(\frac{1}{\sqrt{c+2n}}+\frac{1}{\sqrt{c-2n}}\right) |\nu| + O(\nu^2)\\
    &= \begin{cases}
        2-|\nu|/2 , & \text{Bosonic}\\
        \sqrt{2}-|\nu|/\sqrt{2} , & \text{Type II}\\
        1+\frac{1}{\sqrt{2}}-\frac{1}{4}(1+\sqrt{2})|\nu|,& \text{Heterotic}
    \end{cases} + O(\nu^2).
\end{split}
\end{equation}

In the target space, the winding wavefunction $\chi(\rho)$ is homogeneous in the orthogonal $\vec X$ space, but localised in the rotation plane around $\rho=0$ ($X^1=X^2=0$) with length scales
\begin{equation}\label{eq:winding_mode_length_scale}
    L_{||}^2 \sim \alpha'/|\nu|, \quad L_\perp \sim \infty.
\end{equation}
For the EFT to be self-consistent, we need all the length scales to be much larger than the string scale. Thus, the calculations of this section are strictly valid for $|\nu| \ll 1$. In practice, since we know the physical mass is a finite polynomial in $\nu^2$ (in a given fundamental region), we expect that finitely many $\alpha'$ corrections will give the exact answer as a function of $\nu^2$. However, it is important to keep in mind that the winding EFT is strictly invalid for generic values of $\nu$.

\section{The (slowly) rotating string star}
\label{sec:sss}
In the previous section, we studied properties of free strings. In this section, we turn on the string coupling $g_s$, or the gravitational constant $G_N$. Instead of the free string partition function \eqref{eq:grand_nu}, we will consider the string theory partition function with only asymptotically $\mathbb{R}^{d} \times S^1_R$ geometry (times some compact manifold to saturate the appropriate critical dimension, which we will suppress), twisted by some $\nu \ne 0$ \eqref{eq:WS_ident}. Perturbatively in $G_N$, we can consider different saddles of the partition function with these boundary conditions. One such saddle is the twisted $\mathbb R^d \times S^1_R$ itself, representing weakly-interacting thermal strings in flat space. Another is the (rotating) Euclidean black hole saddle, in which the thermal circle shrinks with an $S^{d-1} \times \mathbb{R}^2$ topology. This saddle remains under control (weakly-curved) only for $\beta \gg l_s$.
At $\nu=0$ and for temperatures close to the Hagdorn temperature, \cite{Horowitz:1997jc} found another type of saddle, a `string star'. Using the EFT found in \cite{Atick:1988si} coupled to gravity, the authors found a normalizable condensate of the winding mode $\chi$ for $3\le d \le 5$ - a string star. In this section, we will find a similar saddle for the case of $\nu \ne 0$, a rotating string star.

Analogously to the $\nu =0$ case, we work at temperatures close to the Hagedorn temperature $R_H(\nu)$. As the winding mode becomes parametrically light compared to the other stringy modes, we can look for saddles of the EFT derived in the previous section, together with gravity.  Analogously to the nonrotating case~\cite{Chen:2021dsw,Horowitz:1997jc}, we will keep only quadratic fluctuations of the metric and a cubic interaction of the metric components with the winding mode, and verify a posteriori that the solution is self-consistent. Due to the more complicated structure of the metric, more metric components couple at this order to the winding mode compared to the $\nu=0$ case, including also the dilaton.

As explained at the end of the last section, for the momenta to be small compared to the string scale, the EFT is valid only for $\nu \ll 1$. After writing down the string star equations of motion, we will take a further simplification and treat $\nu$ perturbatively. We use the perturbative expansion to find numerically the string star profile to first non-trivial order $O(\nu^2)$. Using the expansion, we also write down the thermodynamic properties of the rotating string star, both in the canonical and microcanonical ensembles, which, to first order, amounts to the rigid-body approximation (as we saw in the previous section).

For the metric, it will prove useful to use spherical coordinates for the remaining $d-2$ directions. Starting from \eqref{eq:new_flat_metric_1}, we define $z>0$ as the radius in the remaining $d-2$ directions\footnote{For $d=3$, $z$ takes values in the real axis.}, which leads to
\begin{equation}\label{eq:new_flat_metric}
\begin{split}
    ds^2 &= R^2 dt^2 + d\rho^2 + \rho^2 (d\phi + \nu dt)^2 +  dz^2 + z^2 d\Omega^2_{d-3},\\
    &= R^2 dt^2 + dr^2 + r^2 \left(
    d\theta^2 + \sin^2\theta \ (d\phi - \nu dt)^2
    + \cos^2\theta \ d\Omega^2_{d-3}\right).
\end{split}
\end{equation}
As before, $t\sim t+2\pi$ and $\phi \sim \phi+2\pi$ with no coupling in their identification. $\rho$ and $\phi$ denote the radius and angle in the rotation plane. In the second line, we define the total radius $r^2=\rho^2+z^2$, with the angle $\theta$ such that $\rho = r \sin \theta$ and $z=r \cos \theta$. In other words, $\theta=\pi/2$ is the rotation plane, while $\theta=0$ is the perpendicular $d-2$-dimensional hypersurface.

Finally, we focus in this section only on bosonic and type II strings. The reason is that for heterotic strings the winding mode has non-zero thermal momentum $n$ \eqref{eq:wind_param}, and thus couples in a more complicated way to the deformed metric. We believe that the difference is not essential, and finding solutions for heterotic strings is straightforward.

\subsection{The effective action}\label{subsec:ss_eft}
The low-energy effective action is composed of two parts $I = I_\text{grav}+I_\chi$. $I_\text{grav}$ is the (super)gravity low-energy action of the string theory in question, around $\mathbb R^d \times S^1$. As we are going to excite only the winding mode $\chi$, it is enough to consider the backreaction of the metric and the dilaton~\cite{Horowitz:1997jc}, with the action\footnote{At higher orders in $R-R_H \ll l_s$, the winding mode could also excite a non-zero NS-NS flux.}
\begin{equation}
    I_\text{grav} = \frac{1}{8 G_N} \int d^{d} x \sqrt{G} e^{-2\Phi}  \left( - \mathcal{R} - 4(\partial \Phi)^2 \right),
\end{equation}
with $x$ denoting the $d$ spatial coordinates, and $\sqrt{G}$, $ \mathcal{R}$ are the $d+1$-dimensional volume and Ricci scalar of the deformed metric. We intend to deform $\chi$ while preserving the $SO(2)\times SO(d-2)$ symmetry of the background \eqref{eq:new_flat_metric}. 
The most general deformation of the flat space metric \eqref{eq:new_flat_metric} that preserves this symmetry is given by
\begin{equation}\label{eq:deform_metric}
    ds^2 = R^2 e^{2\varphi} dt^2 + e^{2\eta} d\rho^2 + \rho^2e^{2\alpha} (d\phi - e^{\Psi} \nu dt)^2
    +  dz^2 + e^{2\zeta} z^2 d\Omega^2_{d-3}.
\end{equation}
Due to the $SO(2)\times SO(d-2)$ symmetry, we assume all the fields depend solely on $\rho,z$. In those variables, the gravitational action takes the form
\begin{equation}
    \begin{split}
        I_\text{grav} & = \frac{\Omega_{d-3}}{8 G_N} \int d\rho dz \sqrt{G} e^{-2\Phi}\Bigg[  \frac{\rho^2 \nu^2}{2R^2} e^{-2(\varphi-\alpha-\Psi)} (\nabla \Psi)^2 + 2 \nabla_{\mu}(2\Phi - (d-3)\zeta) \nabla^{\mu}(\alpha + \varphi) \\ & + 2(d-3)\frac{1}{z}\partial_z \eta - 2 \nabla_{\mu} \varphi \nabla^{\mu} \alpha  + 2 (\partial_z \eta)^2 + (\frac{1}{\rho}\partial_{\rho}\alpha - \frac{1}{\rho}\partial_{\rho} \eta) 2 e^{-2\eta} + 2 \partial_z^2 \eta \\ &
        - (d-3)(d-4) (\nabla \zeta)^2 - (d-3)(d-4) \frac{1}{z^2}(e^{-2\zeta}-1) + 2(d-3)\frac{\partial_z \zeta}{z} \\
        &+ 4(d-3) \nabla_{\mu} \zeta \nabla^{\mu} \Phi
        - 4(\nabla \Phi)^2 \Bigg],
    \end{split}
\end{equation}
with $ \sqrt{G}= R \rho z^{d-3} e^{\varphi+(d-3)\zeta+\eta+\alpha}$  and $\Omega_n$ the unit volume of the n-sphere.

The effective theory for the winding mode $\chi$ was described in section \ref{subsec:Atick_W}. Including the coupling to the metric and the dilaton, we rewrite \eqref{eq:eft_chi2} as
\begin{equation}\label{eq:chi_action}
    I_\chi = \frac{1}{8 G_N} \int d^{d} x \sqrt{G} e^{-2\Phi} \left( |\partial \chi|^2 + m^2 |\chi|^2\right).
\end{equation}
With the deformed metric \eqref{eq:deform_metric}, the winding mass 
\eqref{eq:chi_mass} takes the form (for bosonic and type II string theories) 
\begin{equation}
    m^2 = \frac{1}{(\alpha')^2}\left(
    R^2 e^{2\varphi} + \rho^2 \nu^2 e^{2(\alpha + \Psi)} - R_H^2
    \right).
\end{equation}
Here and in the rest of the section, $R_H^2$ will stand for the $\nu=0$ value (or ``$c$'', in the language of \eqref{eq:chi_mass}), while the $\nu$-dependent physical Hagedorn temperature will be denoted as $R_H(\nu)$.
Following the logic of the previous section, we ignore higher $\alpha'$ corrections which are expected to contribute for $\nu \ne 0$ due to couplings to the KK gauge field. Since these will be subleading for any self-consistent solution of the EFT, we will consistently ignore them in this section.

To leading order in $R-R_H \ll l_s$, it is enough to expand the gravitational action $I_\text{grav}$ to quadratic order, and study the cubic interaction of $\chi$ through $I_\chi$~\cite{Chen:2021dsw,Urbach:2022xzw}. 
It is useful to introduce a $d$-dimensional dilaton field
\begin{equation}
    \phi_d = \Phi - \frac{1}{2}(\alpha + \varphi+\eta) - \frac{d-3}{2} \zeta.
\end{equation}
Expanding the gravitational action to quadratic order gives the effective action
\begin{equation}\label{eq:I_2_gravity}
    \begin{split}
        I_\text{grav}^{(2)} &
        = \frac{2\pi R\Omega_{d-3}}{16\pi G_N} \int d\rho \rho dz z^{d-3} \Bigg[ \frac{\rho^2 \nu^2}{2R^2}  (\partial_i \Psi)^2 +  (\partial_i \alpha )^2 + (\partial_i \varphi)^2 + (d-3)(\partial_i \zeta)^2 - 4(\partial_i \phi_d) ^2 \\ & 
        - 4(\partial_{\rho}\phi_d)(\partial_{\rho} \eta) + 4(d-3)\frac{\zeta}{z}\partial_z \phi_d - \frac{4}{\rho} \partial_{\rho}(\alpha-\eta) \phi_d -2 (d-3)(d-4) \frac{\zeta^2}{z^2}  \\
        & - \frac{4}{\rho}(\partial_{\rho} \alpha ) \eta +(\partial_z \eta)^2-(\partial_{\rho}\eta)^2
         \Bigg],
    \end{split}
\end{equation}
where $i \in \{\rho,z\}$.
Expanding the winding mode action to cubic order to leading order in $(R-R_H)/l_s$ leads to 
\begin{equation}\label{eq:chi_action_2}
    I_{\chi} = \frac{2\pi R \Omega_{d-3}}{16\pi G_N} 
    \int d\rho \rho dz z^{d-3} \left[ 
    |\partial_i \chi|^2 + m_0^2 |\chi|^2 + \frac{\nu^2\rho^2}{\alpha'^2} |\chi|^2 +\left( \frac{\kappa}{\alpha'} \varphi + 2\frac{\nu^2 \rho^2}{\alpha'^2} (\alpha+\Psi)\right) |\chi|^2 \right],
\end{equation}
where, following~\cite{Chen:2021dsw}, we have introduced the notation 
\begin{equation}
    m_0^2 = \frac{R^2-R_H^2}{\alpha'^2}, \quad \kappa =\frac{2R^2_H}{\alpha'}.
\end{equation}

Together, \eqref{eq:I_2_gravity} and \eqref{eq:chi_action_2} gives the effective action to leading order in $(R-R_H)/l_s$.
As in the non-rotating case, the non-trivial cubic coupling comes from the mass term.
We see that only $\varphi, \alpha$ and $\Psi$ couple directly to the winding mode. However, due to the (friction) quadratic couplings, also $\phi_d$, $\eta$, and $\zeta$ are turned on. All of the quadratic couplings are of the same order in the characteristic scale $1/m_0$. Notice that for $d=3$, the situation slightly simplifies as $\zeta$ doesn't exist (and $z$ takes real values).

For the EFT to be self-consistent, the winding mode mass needs to stay light compared to the string scale. The effective mass in \eqref{eq:chi_action_2} has two terms. The first requires $\alpha' m_0^2 \ll 1$, or $(R-R_H)/R_H \ll 1$, which we already assumed to expand the action. The second requires $\nu^2 \rho^2 /\alpha' \ll 1$ for the characteristic scale of $\rho$. As we saw in section \ref{subsec:Atick_W}, for the winding mode (which gives the upper bound for the string star size) $\rho^2 \sim \alpha'/|\nu|$, which gives the second condition $\nu^2 \ll 1$. Together, we find that the EFT is valid for
\begin{equation}\label{eq:eft_valid}
    \alpha' m_0^2 \ll 1, \quad \nu^2 \ll 1.
\end{equation}

Finally, the equations of motion from \eqref{eq:I_2_gravity}, \eqref{eq:chi_action_2} are
\begin{equation}\label{eq:full_eoms}
    \begin{split}
        & \frac{1}{\rho z^{d-3}}\partial_{i}(\rho z^{d-3}\partial_{i}\chi) = m_0^2 \chi + \frac{\nu^2 \rho^2 
        }{\alpha'^2} \chi + \left(\frac{\kappa}{\alpha'}\varphi + 2\frac{\nu^2 \rho^2}{\alpha'^2}(\alpha + \Psi)\right)\chi \\
        & \frac{1}{\rho z^{d-3}}\partial_{i}(\rho z^{d-3}\partial_{i}\varphi) = \frac{\kappa}{2\alpha'} |\chi|^2 \\
        & \frac{1}{\rho^3 z^{d-3}}\partial_{i}(\rho^3 z^{d-3}\partial_{i}\Psi) = \frac{\kappa}{\alpha'} |\chi|^2 \\
        & \frac{1}{\rho z^{d-3}}\partial_{i}(\rho z^{d-3}\partial_{i}\alpha) - \frac{2}{\rho}\partial_{\rho}(\phi_d + \eta) = \frac{\nu^2 \rho^2}{\alpha'^2} |\chi|^2 \\
        & \frac{1}{\rho z^{d-3}}\partial_{i}(\rho z^{d-3}\partial_{i}\zeta) = \frac{2}{z}\partial_z \phi_d -2(d-4)\frac{\zeta}{z^2} \\
        & \frac{2}{\rho z^{d-3}}\partial_{i}(\rho z^{d-3}\partial_{i}\phi_d) + \frac{1}{\rho}\partial_{\rho}(\rho \partial_{\rho} \eta) - \frac{d-3}{z^{d-3}} \partial_z(z^{d-4}\zeta) = \frac{1}{\rho}\partial_{\rho}(\alpha-\eta) \\
        & \frac{1}{\rho}\partial_{\rho}(\rho \partial_{\rho} \eta)-\frac{1}{z^{d-3}}\partial_z (z^{d-3} \partial_z \eta) + 2 \partial_{\rho}^2 \phi_d = \frac{2}{\rho}\partial_{\rho} \alpha. 
    \end{split} 
\end{equation}

\subsection{Perturbative solutions}\label{subsec:eom_ss}
We would like to study the string star equations \eqref{eq:full_eoms} in a simplifying limit. It is instructive to first understand the relevant length scales of the equations. At quadratic order, all the fields besides $\chi$ are massless, while $\chi$ has an effective mass term $m_0^2 + \frac{\rho^2\nu^2 }{\alpha'^2}$. We can estimate its effect in various regimes. 

First, arbitrarily close to the modified Hagedorn temperature $(R-R_H(\nu))/l_s \ll \nu \ll 1$, we have $m_0^2\alpha' \sim -|\nu|$. In this regime, we expect the solution profile to be similar to the winding mode solution. Therefore, we expect (see \eqref{eq:winding_mode_length_scale}) $L_{||} \sim l_s/\sqrt{\nu}$ while $L_{\perp} \sim l_s^{\frac{3}{2}}/\sqrt{R-R_H(\nu)}$. In other words, the solution extends mostly in the perpendicular direction and remains localized in the rotation plane \cite{Urbach:2025eyu}.
\footnote{For $d=2$,
the equations for $\varphi$ and $\chi$ are similar to the $AdS_3$ string star equations~\cite{Urbach:2023npi} at leading order in $l_s/l_{ads}$, with $|\nu|=R_H/l_{ads}$.
The analogy to the $AdS_3$ string star equations means that there is a solution to these equations in $d=2$. However, there is no `nonrotating' limit for such a solution, just as there's no flat space limit for the $AdS_3$ string star. See section \ref{sec:ads} for more information about the situation in AdS.} 

For lower temperatures with $(R-R_H(\nu))/l_s \sim |\nu|$, we find from the equations of motion that $\chi,\varphi \sim |\nu|$ with both $L_{||},L_{\perp}\sim l_s /\sqrt{|\nu|}$. However, in this limit, the expansion is not analytic in $\nu$, and there is no good notion of a real rotating solution.

Lastly, we can consider $\nu$ sufficiently small so that we can treat the problem perturbatively in $\nu^2$, and solve the equations order by order. At $\nu=0$, the spherically-symmetric solution has a typical length scale of $L \sim 1/m_0$~\cite{Chen:2021dsw,Horowitz:1997jc}. Since the leading $\nu$ interaction term is controlled by $\nu^2 \rho^2/\alpha'$, demanding it to be perturbative requires that $\nu^2 L^2 \ll R^2-R_H^2$. In terms of $m_0^2$, that means\footnote{Since the EFT is valid only for $\alpha' m_0^2\ll 1$, this also entails $\nu \ll 1$ \eqref{eq:eft_valid}.}
\begin{equation}
    \frac{\nu}{\alpha' m_0^2} \ll 1.
\end{equation}
In this limit, we expect the solution to be almost spherical and to rotate slowly.
In the remainder of the section, we will treat $\nu$ as a perturbation and study the solution to leading non-trivial order in $\nu^2$.

\begin{figure}[h]
    \centering
    \begin{subfigure}{0.3\textwidth}
         \centering
         \includegraphics[width=\linewidth]{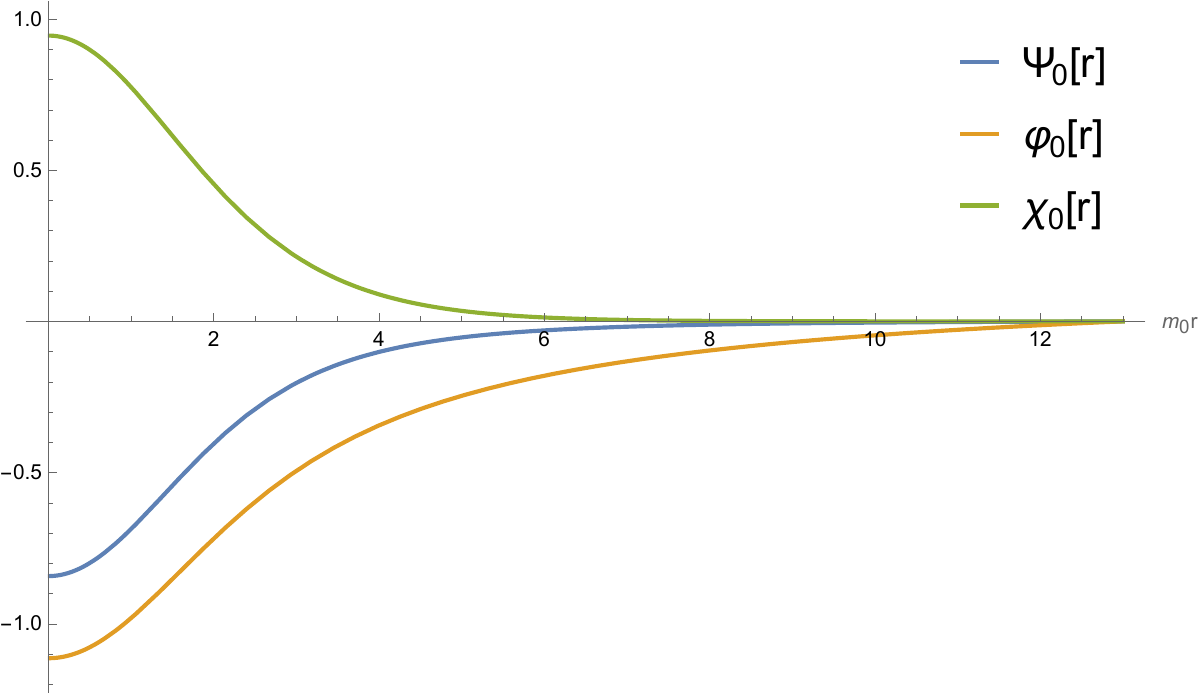}
         \caption{$d=3$.}
         \label{fig:d=3 zeroth order}
     \end{subfigure}
     \begin{subfigure}{0.3\textwidth}
         \centering
         \includegraphics[width=\linewidth]{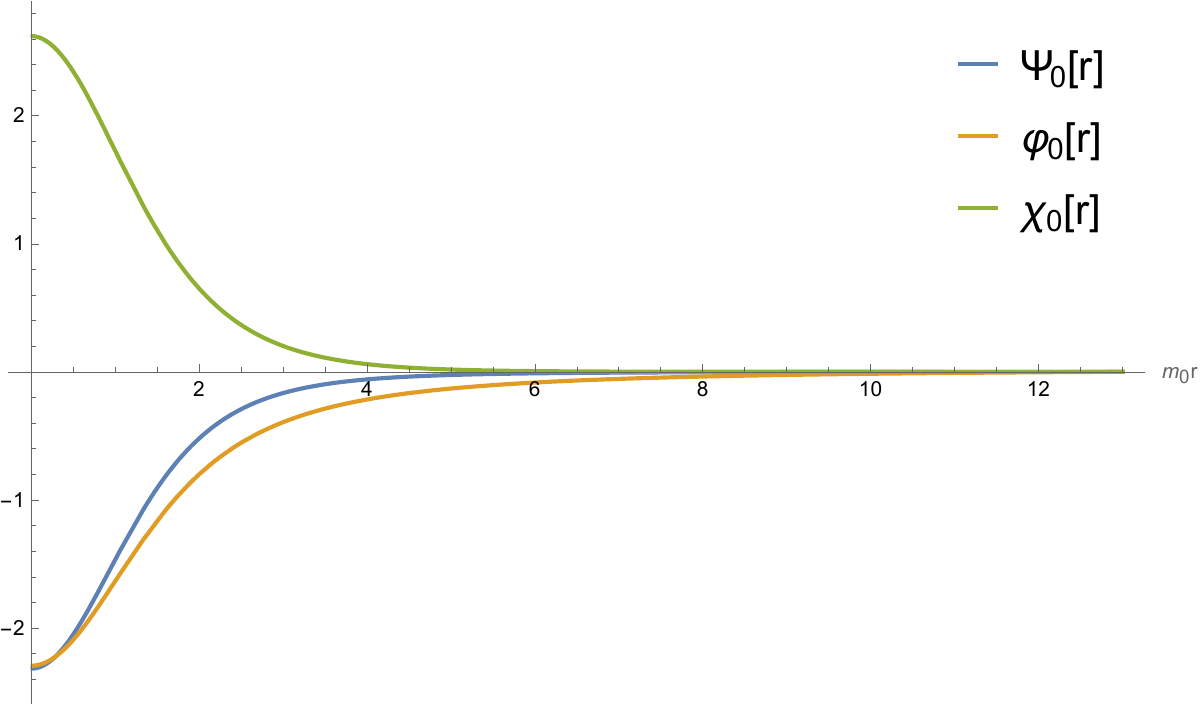}
         \caption{$d=4$.}
         \label{fig:d=4 zeroth order}
     \end{subfigure}
     \begin{subfigure}{0.3\textwidth}
         \centering
         \includegraphics[width=\linewidth]{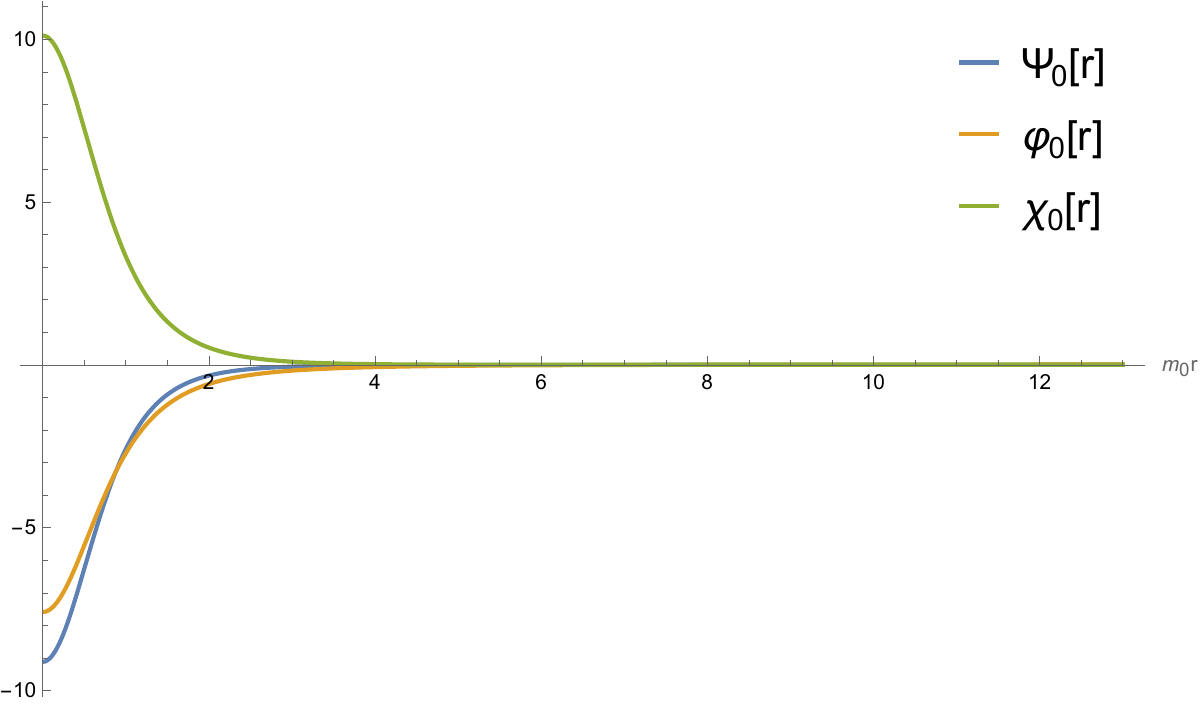}
         \caption{$d=5$.}
         \label{fig:d=5 zeroth order}
     \end{subfigure}
    \caption{The zeroth order solution for $d=3,4,5$ as a function of $m_0 r$. The solutions have been obtained numerically in Mathematica, with a shooting method to get the correct asymptotic behaviour.}
    \label{fig:zeroth order solutions}
\end{figure}

To study the equations, we introduce the following perturbative expansion of the fields
\begin{equation}\label{eq:nu_expansion}
    \begin{split}
        & \chi= \frac{2\alpha' m_0^2}{\kappa}\left(\chi_0 +  \left(\frac{\nu}{\alpha' m_0^2}\right)^2 \chi_1 + \mathcal{O}(\nu^4)\right), \\
        & \varphi= \frac{2\alpha' m_0^2}{\kappa}\left(\varphi_0 +  \left(\frac{\nu}{\alpha' m_0^2}\right)^2 \varphi_1 + \mathcal{O}(\nu^4)\right), \\
        & \Psi= \frac{2\alpha' m_0^2}{\kappa}\left(\Psi_0 +  \left(\frac{\nu}{\alpha' m_0^2}\right)^2 \Psi_1 + \mathcal{O}(\nu^4)\right), \\
        & \alpha = \nu^2 \mathbb \alpha_1 + \mathcal{O}(\nu^4), \quad 
        \phi_d = \nu^2 \mathbb \phi_{d,1} + \mathcal{O}(\nu^4),\\
        & \zeta = \nu^2 \mathbb \zeta_1 + \mathcal{O}(\nu^4),
        \quad 
        \eta = \nu^2 \mathbb \eta_1 + \mathcal{O}(\nu^4).
    \end{split}
\end{equation}
At zeroth order, we simply set $\nu=0$ in the equations. In this limit, it is easy to show that only $\Psi,\varphi, \chi$ admit a nontrivial solution, while the last four equations imply that $\alpha, \phi_d, \zeta, \eta$ are of order $O(\nu^2)$. This explains the leading terms in \eqref{eq:nu_expansion}. 
The overall factor in \eqref{eq:nu_expansion} for $\chi,\varphi,\Psi$ was chosen to put the equation of motion into a scale-independent form.
At zeroth order, the equations take the form
\begin{equation}\label{eq:zeroth_order_r_eq}
    \begin{split}
        & \frac{1}{m_0^2 r^{d-1}}\partial_r\left(r^{d-1} \partial_r \chi_0\right) =  \chi_0 +2\varphi_0 \chi_0  \\
        & \frac{1}{m_0^2 r^{d-1}}\partial_r\left(r^{d-1} \partial_r \varphi_0\right) = \chi_0^2\\
        & \frac{1}{m_0^2 r^{d+1}}\partial_r\left(r^{d+1} \partial_r \Psi_0\right) = 2 \chi_0^2.
    \end{split}
\end{equation}
To write this equation, we moved to the $r, \theta$ variables introduced in \eqref{eq:new_flat_metric}, and took $\chi$ to be real. As these are the $\nu=0$ equations, the first two equations are the standard (non-rotating) Horowitz-Polchinski equations~\cite{Horowitz:1997jc,Chen:2021dsw}. 
By absorbing $m_0$ into a redefinition of $r$, the equations are scale-invariant. As a result, the solution's length scale and amplitude are $\sim 1/m_0$ and $\alpha' m_0^2$ respectively. Together, this justifies the gradient expansion and quadratic approximation we made a posteriori, in the near-Hagedorn temperature limit $m_0^2 \alpha' \ll 1$. The same holds for the novel $\Psi_0$ equation.
Normalizable solutions to the equations \eqref{eq:zeroth_order_r_eq} exist (numerically) for $3\le d \le 5$, see figure \ref{fig:zeroth order solutions}. For $d\geq 6$, there are no self-consistent solutions for the EFT, as the solution will always have an instability towards shrinking down~\cite{Chen:2021dsw,Horowitz:1997jc} (see~\cite{Balthazar:2022szl,Balthazar:2022hno,Bedroya:2024igb,Chu:2025boe,Chu:2025kzl} for recent progress about stringy solutions at $d>5$).

To the first non-trivial order of the fields, we find the equations
\begin{equation}\label{eq:1st_order}
    \begin{split}
        & \frac{1}{m_0^2 z^{d-3}\rho }\partial_i( z^{d-3} \rho \partial_i \chi_1) = \chi_1+ (\rho m_0)^2\left(1 + 4 \frac{\alpha' m_0^2}{\kappa} \Psi_0 \right)\chi_0 
        +2 \varphi_1 \chi_0 +2 \varphi_0 \chi_1  \\
        & \frac{1}{m_0^2 z^{d-3}\rho}\partial_i( z^{d-3} \rho \partial_i \varphi_1) = 2 \chi_0 \chi_1 \\
        & \frac{1}{m_0^2 z^{d-3} \rho^3}\partial_i (\rho^3 z^{d-3} \partial_i \Psi_1) = 4 \chi_0 \chi_1 \\
        & \frac{1}{ m_0^2z^{d-3}\rho} \partial_i (z^{d-3} \rho \partial_i \alpha_1) -\frac{2}{m_0^2 \rho}\partial_{\rho}(\phi_{d,1}+\eta_{1})= 4 \frac{(\rho m_0)^2}{\kappa^2} \chi_0^2 \\
        & \frac{1}{\rho z^{d-3}}\partial_{i}(\rho z^{d-3}\partial_{i}\zeta_1) = \frac{2}{z}\partial_z \phi_{d,1} -2(d-4)\frac{\zeta_1}{z^2} \\
        & \frac{2}{\rho z^{d-3}}\partial_{i}(\rho z^{d-3}\partial_{i}\phi_{d,1}) + \frac{1}{\rho}\partial_{\rho}(\rho \partial_{\rho} \eta_1) - \frac{d-3}{z^{d-3}} \partial_z(z^{d-4}\zeta_1) = \frac{1}{\rho}\partial_{\rho}(\alpha_1-\eta_1) \\
        & \frac{1}{\rho}\partial_{\rho}(\rho \partial_{\rho} \eta_1)-\frac{1}{z^{d-3}}\partial_z (z^{d-3} \partial_z \eta_1) + 2 \partial_{\rho}^2 \phi_{d,1} = \frac{2}{\rho}\partial_{\rho} \alpha_1.
    \end{split}
\end{equation}
Once again, after rescaling $\rho, z$ by $m_0$ and since $\kappa$ is an order one constant, these equations are scale-independent. The only exception is the equation for $\chi_1$. However, since $\alpha' m_0^2 \Psi_0\ll 1$, this term can be dropped consistently.

Notice that the equations are partially decoupled: we can first solve the equations for $\chi_1,\varphi_1$, and then solve the equations for $\Psi_1$, while the equations for $\alpha_1, \phi_{d,1}, \zeta_1,\eta_1$ can be solved independently from those. We focus on $\chi_1,\varphi_1$ to see the corrections to the string star profile and the Newtonian potential.
The first equation has a source term proportional to $\rho^2 = r^2 \sin^2 \theta$. As a result, the solutions are no longer spherically symmetric; instead, there is a monopole and a quadrupole contribution. Indeed, for general $d$, it is easy to see that the generalized spherical harmonic 
\begin{equation}
    Y_{2,0,\dots,0} \propto \sin^2 \theta - \frac{2}{d}, 
\end{equation}
and threfore $\sin^2 \theta$ is a combination of $Y_{2,0,\dots,0}$ and $Y_{0,0,\dots,0}$.\footnote{The kinetic term for $\Psi$ is slightly different: it has a $\rho^3$ instead of a $\rho$ inside the derivative. There, the same expressions hold, but with $d\rightarrow d + 2$.} The spin zero mode corresponds to a small change of the overall radial profile, while the spin two contribution squashes the string star profile as a function of $\theta$.

Decomposing $\chi_1$ and $\varphi_1$ into a monopole and quadrupole moment, we write
\begin{equation}\label{eq:ylm_expansion}
    \begin{split}
        & \chi_1 = \frac{2}{d} \chi_{1,0} + \left(\sin^2 \theta - \frac{2}{d}\right) \chi_{1,2} \\
        & \varphi_1 = \frac{2}{d} \varphi_{1,0} + \left(\sin^2 \theta - \frac{2}{d}\right) \varphi_{1,2},
    \end{split}
\end{equation}
such that the equations governing the deformation of the string star are
\begin{equation}\label{eq:mon_quad_eqs}
    \begin{split}
        & \frac{1}{m_0^2}\left(\partial_r^2 + \frac{d-1}{r}\partial_r\right) \chi_{1,0} = \chi_{1,0} + (m_0r)^2 \chi_0 + 2 \chi_0 \varphi_{1,0} + 2\varphi_0 \chi_{1,0} \\
        & \frac{1}{m_0^2}\left(\partial_r^2 + \frac{d-1}{r}\partial_r - \frac{2d}{r^2}\right) \chi_{1,2} = \chi_{1,2} + (m_0r)^2 \chi_0 + 2 \chi_0 \varphi_{1,2} + 2\varphi_0 \chi_{1,2} \\
        & \frac{1}{m_0^2} \left( \partial_r^2 + \frac{d-1}{r} \partial_r \right) \varphi_{1,0} = 2\chi_0 \chi_{1,0} \\
        & \frac{1}{m_0^2} \left( \partial_r^2 + \frac{d-1}{r} \partial_r - \frac{2d}{r^2} \right) \varphi_{1,2} = 2\chi_0 \chi_{1,2}.
    \end{split}
\end{equation}

\begin{figure}
     \centering
     \begin{subfigure}{0.3\textwidth}
         \centering
         \includegraphics[width=\linewidth]{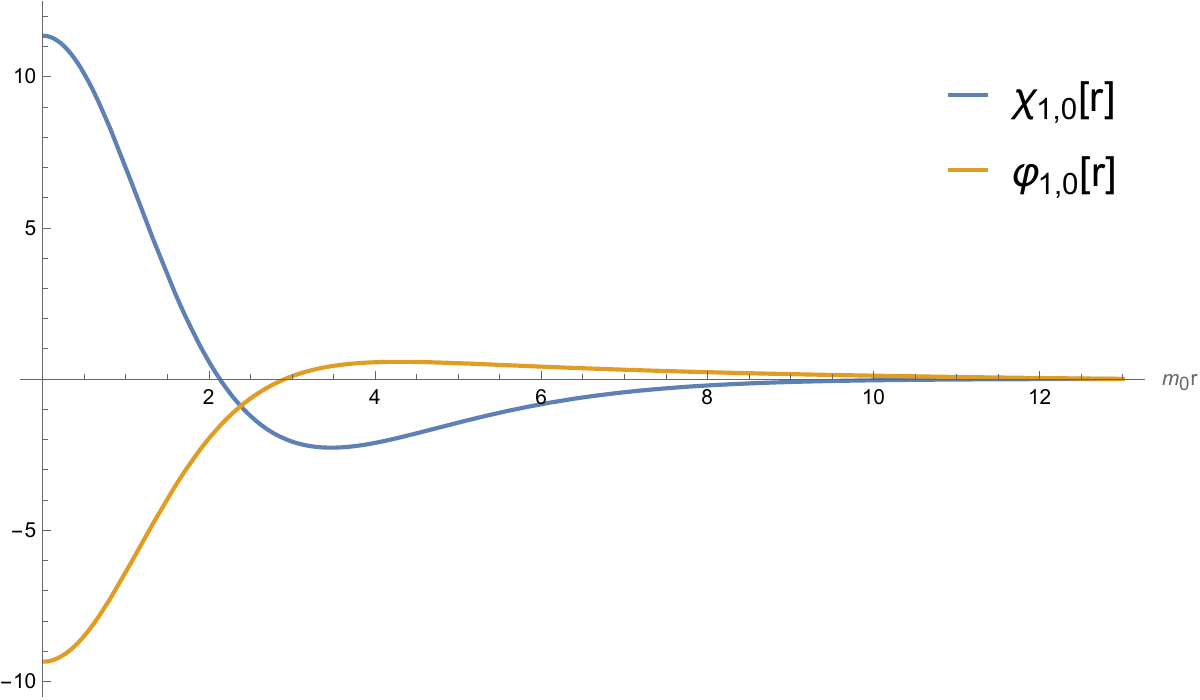}
         \caption{$d=3$ monopole correction.}
         \label{fig:1a}
     \end{subfigure}
     \begin{subfigure}{0.3\textwidth}
         \centering
         \includegraphics[width=\linewidth]{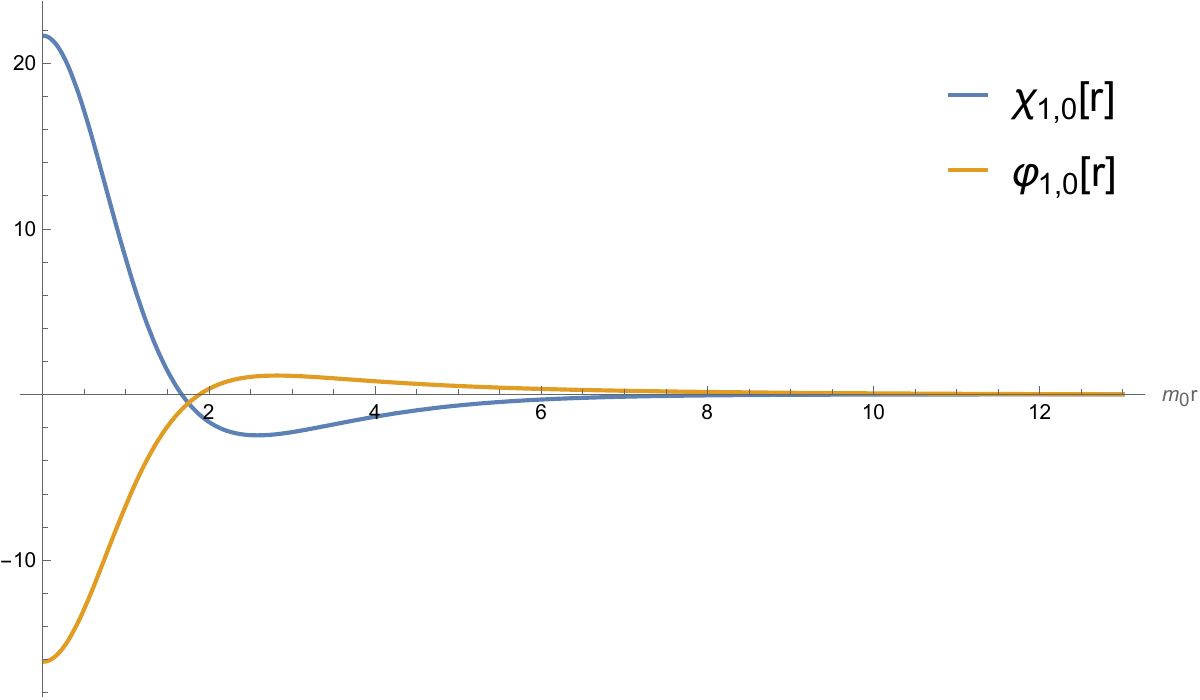}
         \caption{$d=4$ monopole correction.}
         \label{fig:1b}
     \end{subfigure}
     \begin{subfigure}{0.3\textwidth}
         \centering
         \includegraphics[width=\linewidth]{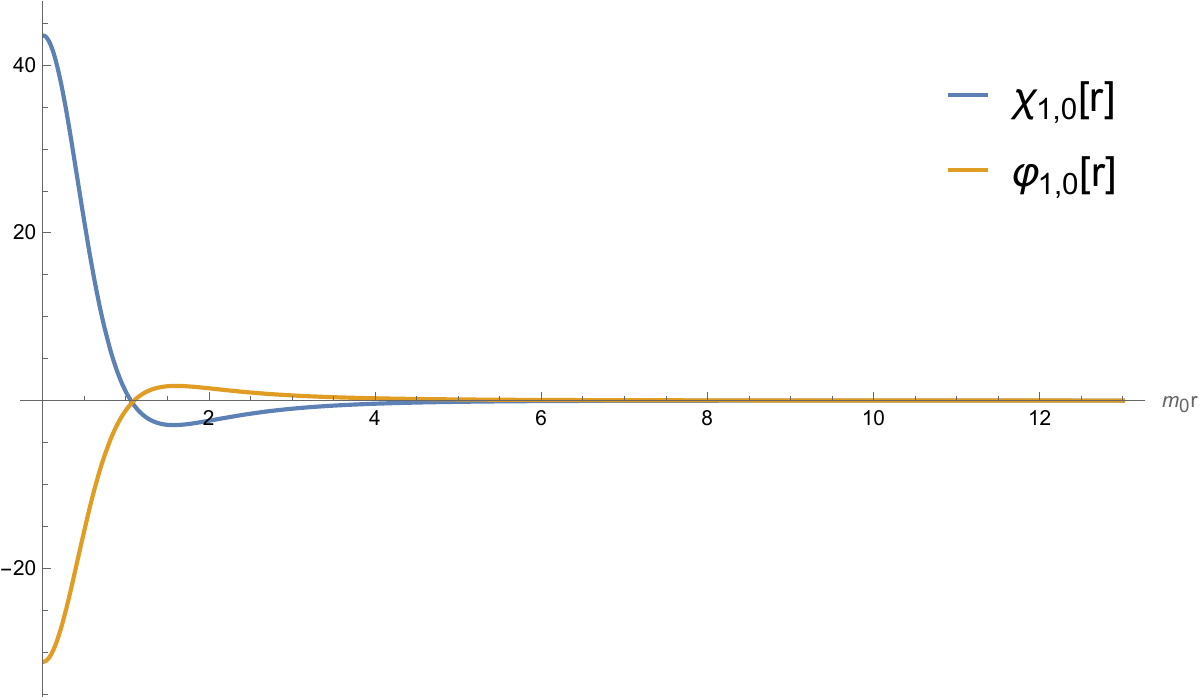}
         \caption{$d=5$ monopole correction.}
         \label{fig:1c}
     \end{subfigure}

     \bigskip
     \begin{subfigure}{0.3\textwidth}
         \centering
         \includegraphics[width=\linewidth]{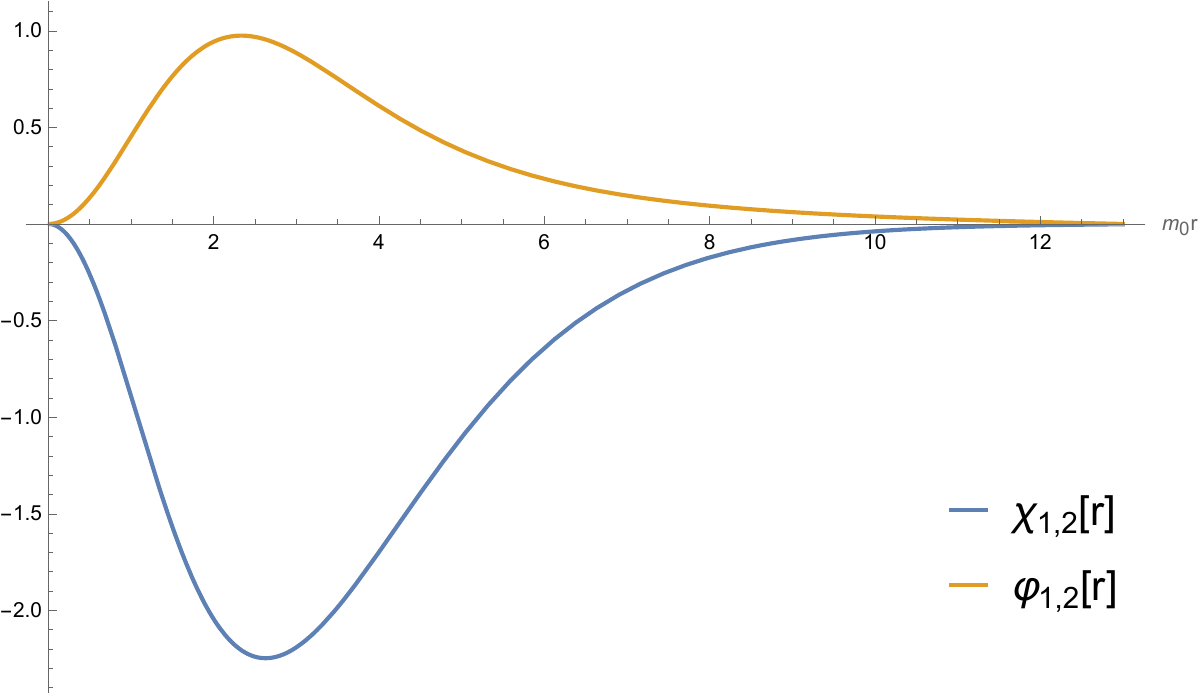}
         \caption{$d=3$ quadrupole correction.}
         \label{fig:1d}
     \end{subfigure}
     \begin{subfigure}{0.3\textwidth}
         \centering
         \includegraphics[width=\linewidth]{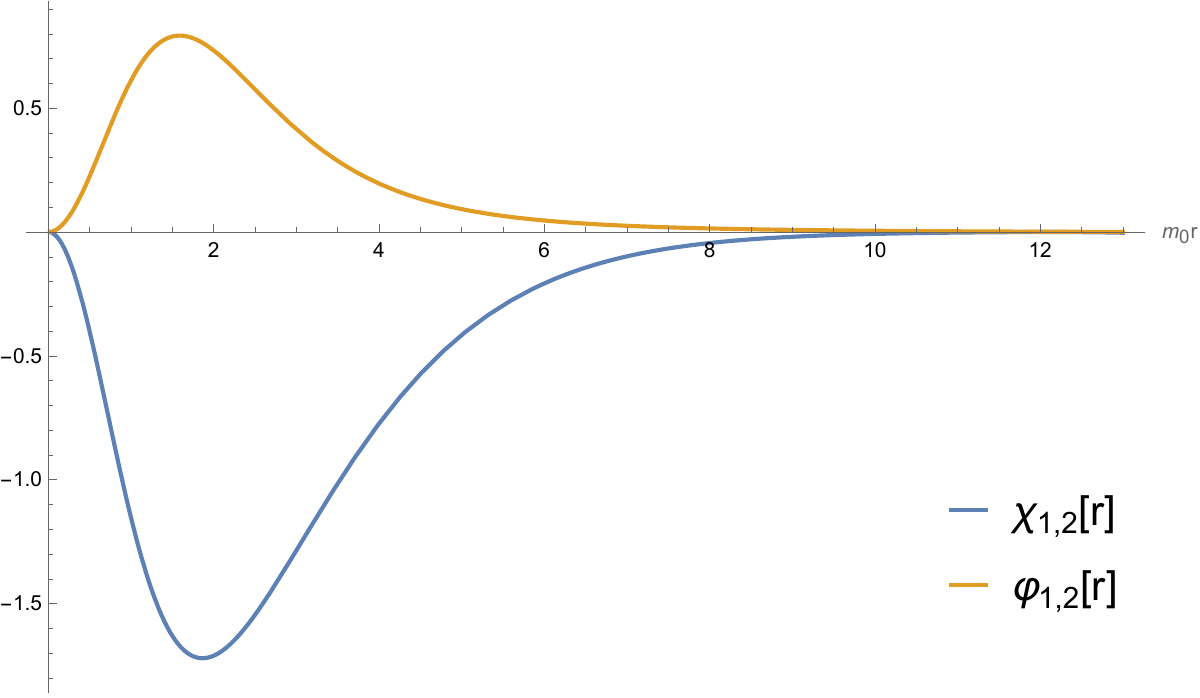}
         \caption{$d=4$ quadrupole correction.}
         \label{fig:1e}
     \end{subfigure}
     \begin{subfigure}{0.3\textwidth}
         \centering
         \includegraphics[width=\linewidth]{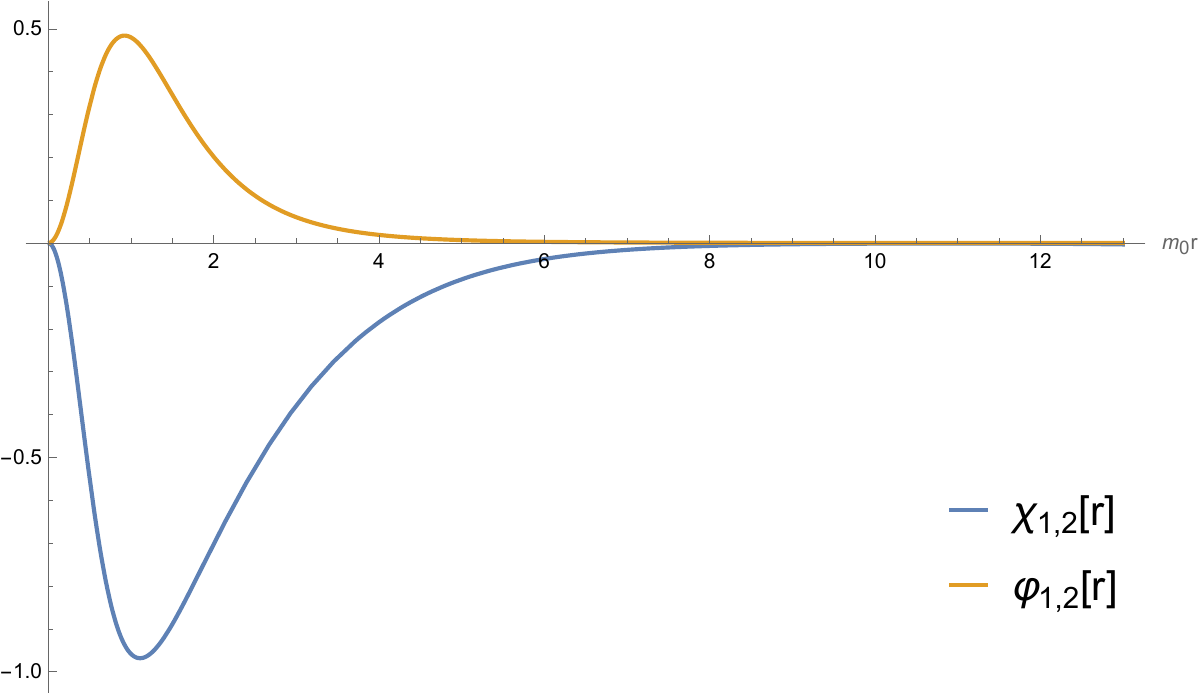}
         \caption{$d=5$ quadrupole correction.}
         \label{fig:1f}
     \end{subfigure}
     \caption{The first order corrections to winding mode and the gravitational potential profiles for $d=3,4,5$, as a function of $m_0 r$. The solutions have been obtained numerically by solving equations \eqref{eq:mon_quad_eqs} using a shooting method.}
     \label{fig:first order corrections}
\end{figure}

It is straightforward to find numerical solutions to \eqref{eq:mon_quad_eqs} using the shooting method. The solutions for $d=3,4,5$ are plotted in figure \ref{fig:first order corrections}. As seen in the plots, at the center, the monopole profile of the winding mode is positive and the Newtonian potential negative. Away from the center, both change sign. For the quadropole correction, we find that the $\chi$ correction is always negative and the $\varphi$ correction is positive. Using \eqref{eq:ylm_expansion}, the effect of the quadrupole correction is to increase the string star density near the poles $\theta \sim 0$ and decrease it near the equator $\theta\sim \pi/2$.

\begin{figure}[h]
    \centering
    \includegraphics[width=0.8\linewidth]{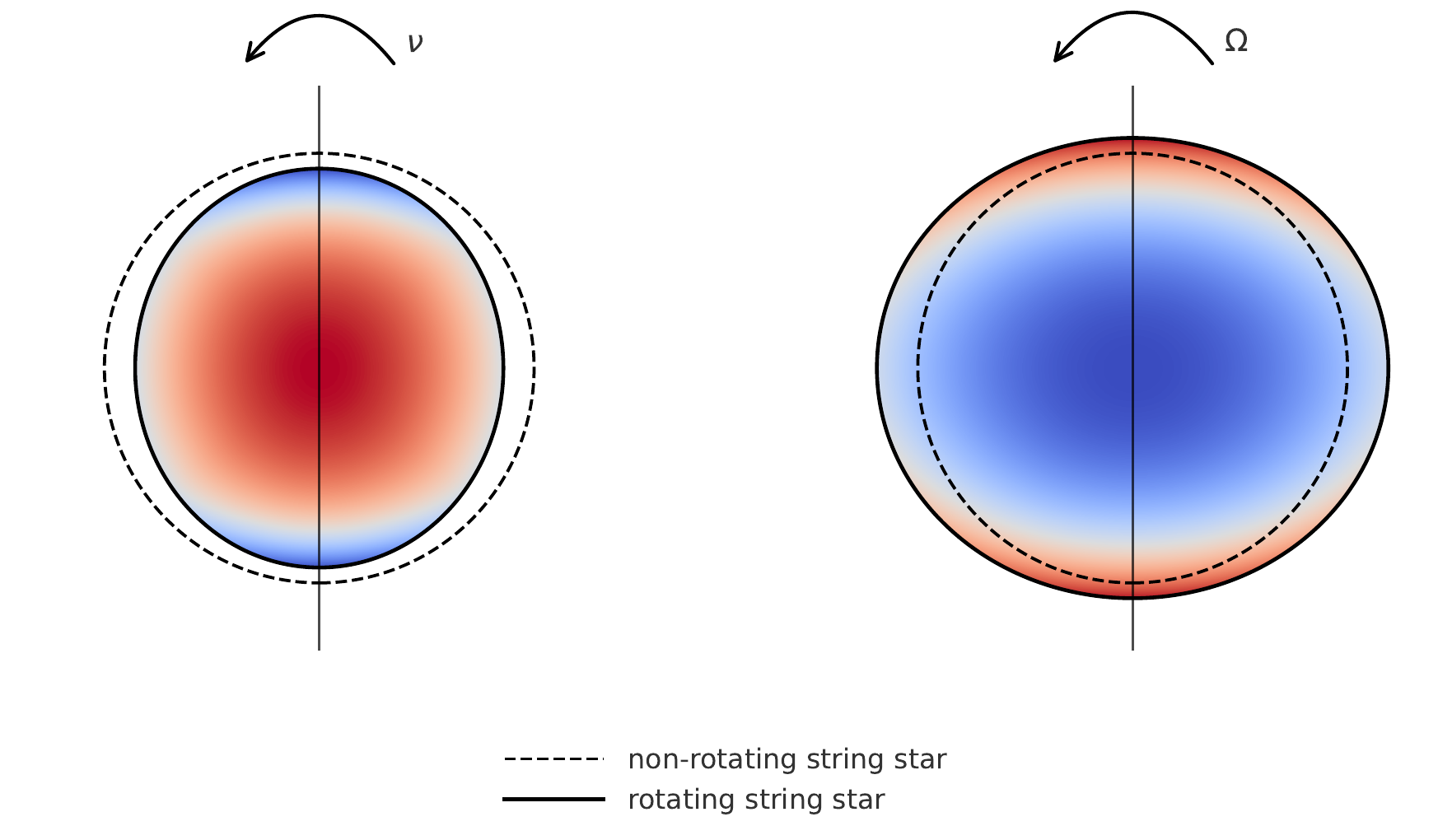}
    \caption{A schematic drawing of the effect of rotation, both in the $\nu$ (left) and $\Omega$ (right) ensembles. Depicted is the cross-section of the string star, with the vertical direction corresponding to $z$ and the horizontal direction to $\rho$. Red areas are of increased density, while blue areas are less dense compared to the nonrotating solution. For real $\Omega$, the string star becomes oblate, as intuitively expected.}
    \label{fig:schematic of rotation}
\end{figure}

We can understand these results using Newtonian intuition. Consider a gravitationally bound spherical gas and let it slowly rotate. Inside the gas, the centrifugal force will push matter outwards, decreasing the density in the center and increasing it towards the surface. Additionally, the rotation will make the sphere slightly oblate, increasing its equator and decreasing its radius at the poles. The Newtonian potential follows in the same way: it is slightly stronger towards the surface and near the equator region due to the higher matter density there, and weaker at the poles and close to the center (see figure \ref{fig:schematic of rotation}). 
Both of these effects are $\mathcal{O}(\Omega^2)$ in the angular velocity. When analytically continuing $\Omega = i \nu/R$, the sign of these effects is reversed: the density near the center increases, and the equator radius decreases, exactly what we find for the rotating string star. 

We would like to estimate the size and shape of the solution in the parallel and perpendicular directions.
We can estimate the sizes as the distance at which the profile has decayed to $1/e$ of its maximal $r=0$ value. By \eqref{eq:nu_expansion} and \eqref{eq:ylm_expansion}, we know that the equator ($L_{||}$) and pole radius ($L_{\perp}$) have the form
\begin{equation}\label{eq:ss_radii}
    \begin{split}
        & L_{||} =  \frac{l_0}{m_0}\left( 1 -c_{||} 
        \frac{\nu^2}{(\alpha' m_0^2)^2} + O(\nu^4)\right) \\
        & L_{\perp} = \frac{l_0}{m_0}\left( 1 -c_{\perp} 
        \frac{\nu^2}{(\alpha' m_0^2)^2} + O(\nu^4)\right), 
    \end{split}
\end{equation}
where $c_{\perp}< c_{||}$ due to the previously described squashing of the string star. Since the $\nu^2$ correction for the mass is nonnegative, the $1/e$ length decreases. The approximate numerical values for $d=3,4,5$ are listed in Table \ref{tab:grand canonical length values}. 
\begin{table}[h]
    \centering
    \begin{tabular}{|c|c|c|c|}
        \hline
        $d$ & $l_0$ & $c_{||}$ & $c_{\perp}$ \\
        \hline \hline 
        3 & 2.38 & 6.76 & 3.12 \\
        4 & 1.64 & 2.88 & 1.84 \\
        5 & 0.939 & 0.985 & 0.830 \\
        \hline
    \end{tabular}
    \caption{Approximate numerical values for the size of the string star. The parameters follow from analyzing the $1/e$ length of the numerical solutions.}
    \label{tab:grand canonical length values}
\end{table}

\noindent
Finally, it is instructive to also write down the eccentricity of the solution, as it has a leading $\nu$ effect. In terms of \eqref{eq:ss_radii}, we find to leading order
\begin{equation}\label{eq:ecc}
    e = \sqrt{1-\frac{L_{\perp}^2}{L_{||}^2}} = \sqrt{2(c_\perp-c_{||})}\frac{|\nu|}{\alpha'm_0^2} + O(\nu^3).
\end{equation}

\subsection{Grand canonical properties}\label{subsec:grand_ss}

We now turn to study the thermodynamic properties of the perturbative rotating string star to leading non-trivial order in $\nu/(\alpha' m_0^2) \ll 1$. Recall that we understand this solution as a (metastable) saddle to the grand canonical partition function $Z(\beta,\nu)$ \eqref{eq:grand_nu} (in what follows, we will use $R$ and $\beta = 2\pi R$ interchangeably). Therefore, our starting point would be to compute the grand canonical on-shell action of the solution. In the perturbative regime, we write the action as an undeformed action plus a perturbation of order $\nu^2$. Therefore, to order $\nu^2$ it is enough to substitute the non-rotating solutions $\chi_0$, $\varphi_0$ and $\Psi_0$ \eqref{eq:zeroth_order_r_eq} in the action. In the Newtonian limit, this is nothing but the standard rigid-body approximation. The potential $\Psi_0$ can be understood as a post-Newtonian effect, and we therefore expect it to be subleading compared to $\varphi_0, \chi_0$. Using the equations of motion, the $\mathcal{O}(\nu^2)$ on-shell action at leading order in $R-R_H$ is
\begin{equation}\label{eq:ss_onshelll_action}
    \begin{split}
        I(R,\nu) & =\frac{(\alpha')^4 R_H^{-3} m_0^{6-d}}{8G_N}\left(\xi_1 + \left(\frac{\nu}{\alpha' m_0^2}\right)^2\  \xi_2 + O\left(\left(\frac{\nu}{\alpha' m^2_0}\right)^4\right)\right),
    \end{split}
\end{equation}
with the two dimensionless $O(1)$ quantities, $\tilde x = m_0 x$,
\begin{equation}\label{eq:xis}
    \begin{split}
        \xi_1 & = \int d^d \tilde x \ (-\varphi_0)\chi_0^2, \\
        \xi_2 & = \int d^d \tilde x \ \chi_0^2 \ \tilde \rho^2.
    \end{split}
\end{equation}
To get the $\xi_2$ expression, we dropped some $\mathcal{O}(\nu^2)$ terms that are subleading in $\alpha' m_0^2$. Since $\xi_1$, $\xi_2$ are both positive $O(1)$ numbers, \eqref{eq:xis} is enough to read off the scaling properties of the rotating string star thermodynamics.

Since we work perturbatively in small angular velocity, we are allowed to rotate back to real angular velocity $\nu = -i R \Omega$. As long as $\Omega$ is small enough, we expect the rotating string star to be a meta-stable solution, with a decay rate non-perturbative in $\Omega$.\footnote{It is also possible to view the $\Omega$-dependent solution as a contribution to the fixed $J$ ensemble $Z(\beta,J)$ (or $Z(M,J)$). In Euclidean gravity, the fixed $J$ ensemble allows the asymptotic $G_{t,\phi} \sim i \Omega \rho^2$ to have any $\Omega$, while the spin is fixed in terms of a subleading mode of $G_{t,\phi}$ (analogous to the microcanonical ensemble where the temperature fluctuate at fixed energy).}
Upon analytic continuation,
\begin{equation}\label{eq:Omega_onshell_action}
    I(R,\Omega) =\frac{(\alpha')^4 R_H^{-3} m_0^{6-d}}{8G_N}\left(\xi_1 - \left(\frac{R_H \Omega}{\alpha' m_0^2}\right)^2\  \xi_2 +...\right).
\end{equation}
By standard thermodynamics, we can now find the thermodynamic properties of the rotating string star.
We begin with the grand canonical mass, given by (to leading order in $m_0$)
\begin{equation}\label{eq:ss_mass}
\begin{split}
    M(R,\Omega) &= (\partial_\beta-\beta^{-1}\Omega \partial_\Omega) I \\
    &= \frac{R_H^2}{8\pi G_N (\alpha')^2} \int d^d x |\chi|^2\\
    &= \frac{(\alpha')^2 R_H^{-2} m_0^{4-d}}{8\pi G_N } \left(  \left(3-\frac{d}{2}\right) \xi_1 + \left(\frac{R_H \Omega}{\alpha' m_0^2}\right)^2 \left(\frac{d}{2}-1\right) \xi_2\right).
\end{split}
\end{equation}
In the second line, following~\cite{Chen:2021dsw}, we used the explicit $\beta$ dependence of the mass and the equations of motion to write the mass as a local integral. In the third line, we took the direct derivative of the action \eqref{eq:Omega_onshell_action}. One can check explicitly that the two agree by the equations of motion. Notice that since $2<d<6$, both terms in the third line are positive.

To leading order in $m_0$, the entropy is
\begin{equation}\label{eq:ss_entropy}
\begin{split}
    S(R,\Omega) &= \left(\beta \partial_\beta - \Omega\partial_\Omega -1\right) I = \frac{(\alpha')^2 R_H^{-1} m_0^{4-d}}{4 G_N } \left( \xi_1 \left(3-\frac{d}{2}\right) + \left(\frac{R_H \Omega}{\alpha' m_0^2}\right)^2 \xi_2 \left(\frac{d}{2}-1\right)\right)\\
    &= \beta_H M.
\end{split}
\end{equation}
The Hagedorn relation $S = \beta_H M$ holds up to subleading corrections in $\beta-\beta_H$, for both the $\mathcal{O}(\Omega^0)$ and $\mathcal{O}(\Omega^2)$ piece.\footnote{This might seem puzzling given the discussion in section \ref{sec:flat_space_therm}, but recall that here we take a limit of very small $\nu$: $\nu \ll \beta-\beta_H$. } We will compute the leading correction to the Hagedorn behaviour in section \ref{subsec:micro_ss}.

Finally, the angular momentum and the moment of inertia are given by 
\begin{equation}\label{eq:ss_J_I}
\begin{split}
    J(R,\Omega) & = -\beta^{-1} \partial_\Omega I = \frac{(\alpha')^2 R_H^{-2} m_0^{2-d}}{8\pi G_N} \xi_2 \cdot \Omega,\\
    \mathcal{I}(R,\Omega) &= -\beta^{-1} \partial_\Omega^2 I = \frac{(\alpha')^2 R_H^{-2} m_0^{2-d}}{8\pi G_N} \xi_2.
\end{split}
\end{equation}
In terms of the moment of inertia $\mathcal{I}$, we can rewrite the angular momentum as $J = \mathcal{I} \cdot \Omega$.
The moment of inertia is also related to the variance of the angular momentum by the standard fluctuation–dissipation relation
\begin{equation}
    \langle \Delta J^2 \rangle = \beta^{-2}\partial^2_\Omega I = \beta^{-1 } \mathcal{I}.
\end{equation}
At the order we are working at, $\Omega$ appears in the action only in the $\rho^2 |\chi|^2$ mass term. As a result, we can use the equations of motion to write both as local expressions. Reading the mass density $\rho(x) = \frac{R_H^2}{8\pi G_N (\alpha')^2} |\chi(x)|^2$ from \eqref{eq:ss_mass}, we find
\begin{equation}\label{eq:ss_J_I_2}
\begin{split}
    J &= \int d^d x \ \Omega\cdot \rho(x)  r^2 \sin^2\theta\\
    \mathcal{I} &= \int d^d x\ \rho(x) r^2 \sin^2\theta,
\end{split}
\end{equation}
precisely following the Newtonian expectation for a rigid mass density $\rho(x)$.

\subsection{Microcanonical properties}\label{subsec:micro_ss}
It is revealing to express the thermodynamic quantities using microcanonical variables. In the previous section, we expanded to leading order in $m_0$ and first order in $(R_H \Omega/(\alpha' m_0^2))^2$. Rewriting \eqref{eq:ss_mass} and \eqref{eq:ss_J_I} gives
    \begin{equation}\label{eq:ss_omega_micro}
    \frac{R_H \Omega}{\alpha' m_0^2} = \frac{R_H}{l_s} \left(3-\frac{d}{2}\right) \frac{\xi_1}{\xi_2}\cdot \frac{J}{l_s M} + ....
\end{equation}
Therefore, in terms of $M,J$ we will expand to leading order in $l_s M \gg 1$ and to first order in $J/(l_s M) \ll 1$. 
Rewriting \eqref{eq:ss_mass} at this order gives
\begin{equation}
    M = \frac{(\alpha')^2 \left(3-\frac{d}{2}\right)\xi_1}{8\pi G_N R_H^{2}} m_0^{4-d} \left(1+
    \left(\frac{d}{2}-1\right)\left(3-\frac{d}{2}\right)\frac{\xi_1}{\xi_2} \left(\frac{R_H}{l_s}\right)^2 \left(\frac{J}{l_s M}\right)^2
    \right),
\end{equation}
which we can invert to find the temperature (using $\alpha' m_0^2 \approx \kappa \frac{R-R_H}{R_H}$)
\begin{equation}\label{temp in micro quantities}
    \frac{(\beta-\beta_H)(M,J)}{\beta_H} = \frac{\alpha'}{\kappa}\left(
    \frac{8\pi G_N M R_H^{2}}{(\alpha')^2 \left(3-\frac{d}{2}\right)\xi_1}\right)^{\frac{1}{2-\frac{d}{2}}} \left(1-
    \frac{\left(\frac{d}{2}-1\right)\left(3-\frac{d}{2}\right)}{2-\frac{d}{2}}\frac{\xi_1}{\xi_2} \left(\frac{R_H}{l_s}\right)^2 \left(\frac{J}{l_s M}\right)^2
    \right).
\end{equation}
To get the microcanonical entropy, we exploit the thermodynamic relation
\begin{equation}
    \left(\frac{\partial S}{\partial M}\right)_J = \beta = \beta-\beta_H+\beta_H,
\end{equation}
and integrate, where we express $\beta-\beta_H$ as a function of $M,J$ via \eqref{temp in micro quantities}. This gives
\begin{equation}\label{eq:ss_entropy_micro}
    S(M,J) = \beta_H M + \frac{\pi \alpha'^2}{R_H} \left(\frac{8\pi G_N M^{3-\frac{d}{2}} R_H^2}{(3-\frac{d}{2})\xi_1 \alpha'^2}\right)^{\frac{1}{2-\frac{d}{2}}}\Bigg[\frac{2-\frac{d}{2}}{3-\frac{d}{2}} - \left(\frac{R_H}{l_s}\right)^2 
    \left(\frac{J}{l_s M}\right)^2 \frac{(3-\frac{d}{2})\xi_1}{\xi_2}\Bigg].
\end{equation}
The leading piece is just the Hagedorn behaviour. The other $J$-independent contribution is the leading correction due to gravitational self-interaction discussed by~\cite{Chen:2021dsw}, while the $J$-dependent piece is the leading correction due to angular momentum. Note that the latter is always negative: some of the energy has to be invested into (entropically suppressed) angular momentum. 

In terms of $M,J$ the sizes \eqref{eq:ss_radii} are
\begin{equation}\label{eq:ss_sizes_micro}
    \begin{split}
        & L_{||}(M,J) = \tilde{l}_0 \left(\frac{G_N  M R_H^2}{\alpha'^2}\right)^{\frac{1}{d-4}} \Bigg[1 + \tilde{c}_{||} \left(\frac{R_H}{l_s}\right)^2\frac{J^2}{(l_sM)^2}\Bigg], \\
        & L_{\perp}(M,J) = \tilde{l}_0 \left(\frac{G_N  M R_H^2}{\alpha'^2}\right)^{\frac{1}{d-4}}\Bigg[1+\tilde{c}_{\perp} \left(\frac{R_H}{l_s}\right)^2 \frac{J^2}{(l_sM)^2}\Bigg],
    \end{split}
\end{equation}
where $\tilde{l}_0$, $\tilde{c}_{||}$ and $\tilde{c}_{\perp}$ are numerical coefficients. Their values for $d=3,5$ are listed in Table \ref{tab:microcanonical length values}.\footnote{We don't list $d=4$ as in that case $\beta-\beta_H$ is to leading order independent of $M$, which means that we need to take the next order into account to go to the microcanonical ensemble~\cite{Chen:2021dsw}.}
For the eccentricity \eqref{eq:ecc} we find
\begin{equation}\label{eq:ecc_micro}
    e = \sqrt{\kappa(c_{||}-c_\perp)} \left(3-\frac{d}{2}\right) \frac{R_H}{l_s} \frac{\xi_1}{\xi_2} \cdot \frac{J}{l_s M}.
\end{equation}
Since $c_{||}>c_\perp$, the solution always turns more oblate as we increase $J$, as dictated by the centrifugal force (see figure \ref{fig:schematic of rotation}).

\begin{table}[h]
    \centering
    \begin{tabular}{|c|c|c|c|c|c|}
        \hline
        $d$ & $\xi_1$ & $\xi_2$ & $\tilde{l}_0$ & $\tilde{c}_{||}$ & $\tilde{c}_{\perp}$\\
        \hline \hline 
        3 & 18.1 & 89.8 & 1.08 & 0.769 & 0.436 \\
        4 & 155 & 313 & - &  - & - \\
        5 & 826 & 394 & 0.06 & -0.490 & -0.660\\
        \hline
    \end{tabular}
    \caption{Approximate numerical parameters for the string star, in the microcanonical ensemble. Note that for $d=5$, the leading $J^2$ correction causes a shrinking of the solution.}
    \label{tab:microcanonical length values}
\end{table}

\section{The correspondence principle}
\label{sec:corr}

The main importance of the string star saddle was the suggestion by Horowitz and Polchinski~\cite{Horowitz:1997jc} that it interpolates between the black hole and the free string phases of string theory. More precisely~\cite{Chen:2021dsw}, the suggestion is that as a Euclidean saddle defined close to $\beta \sim \beta_H$, it is continuously connected to the Euclidean black hole at lower temperatures $\beta \gg \beta_H$. In other words, as the Euclidean black hole size reaches the string scale, its Einstein gravity description is no longer valid, and should now be described as a self-gravitating string.
In this context, this is what is usually referred to as the correspondence principle, or the black hole-string transition. The main argument in favor of this suggestion was that extrapolating the thermodynamic properties of both saddles to an intermediate regime $\beta -\beta_H \sim l_s$  (where both are ill defined), gives qualitatively the same result.

In the other direction, close enough to the Hagedorn temperature, the string star amplitude goes to zero, and we expect it to reduce to the free string phase of the thermal background. This is better described in terms of the microcanonical ensemble. The free string phase is expected to dominate the ensemble at low energies. For high enough energies, the free string self-gravitation can no longer be ignored, and we expect a transition to the string star description. Here as well, a simple extrapolation qualitatively gives the same answer.

In the previous section, we derived the properties of the rotating string star saddle to leading non-trivial order in the angular velocity $O(\Omega^2)$ (or $J^2$ in the microcanonical ensemble). 
In this section, we discuss the extension of the correspondence arguments to the rotating string star saddle, at leading order in $O(\Omega^2)$. This requires a comparison to rotating black holes on one side and rotating free strings on the other. As we will see, the qualitative arguments seem to hold in a similar manner also for the (slowly) rotating case.

\subsection{Rotating black holes}
\label{subsec:black hole string star corr}

We begin by discussing the properties of slowly rotating black holes, following the conventions of~\cite{Emparan:2003sy,Ceplak:2023afb}.
The thermodynamic properties of rotating black holes in $D=d+1$ dimensions (also known as Myers-Perry black holes~\cite{Myers:1986un}) are best described using the variables $\mu$, $a$, and the black hole horizon radius $r_h$. These variables are related by the condition
\begin{equation}
    r_h^2+a^2 = \frac{\mu}{r_h^{d-4}}.
\end{equation}
The mass, entropy, and angular momentum are parametrized in these variables by
\begin{equation}
\begin{split}
    M = \frac{(d-1)\Omega_{d-1}}{16\pi G_N} \mu, \quad S = \frac{\Omega_{d-1}}{4 G_N} r_h^{d-3} (r_h^2+a^2), \quad
    J = \frac{2}{d-1} M a.
\end{split}
\end{equation}
The temperature and angular velocity, on the other hand, are parametrized as
\begin{equation}
    T = \frac{1}{4\pi}\left(\frac{2r_h^{d-3}}{\mu}+\frac{d-4}{r_h}\right),\quad \Omega = \frac{a}{r_h^2+a^2}.
\end{equation}
We can invert the latter relation for small $\Omega$, which gives to first non-trivial order 
\begin{equation}
\begin{split}
    M &= \alpha_d\frac{\beta^{d-2}}{G_N}\left(1-\frac{(d-2)^2}{16\pi^2} \beta^2\Omega^2 + O(\Omega^4)\right)\\
    S &= \frac{d-2}{d-1}\alpha_d\ \frac{\beta^{d-1}}{G_N}\left(1-\frac{d(d-2) }{16\pi^2} \beta^2\Omega^2 + O(\Omega^4)\right)\\
    J &= \frac{(d-2)^2}{d-1}\alpha_d\ \frac{\beta^d}{8\pi^2 G_N} \Omega + O(\Omega^3) ,
\end{split}
\end{equation}
with $\alpha_d = \frac{(d-1)(d-2)^{d-2}\Omega_{d-1}}{4^d \pi^{d-1}}$, and $\beta = T^{-1}$. In all the formulas, the leading term corresponds to the non-rotating Schwarzschild black hole, and the second term is the leading correction due to a small angular velocity $\Omega$.

The overall size of the black hole is captured by its horizon radius, which to leading order is
\begin{equation}\label{eq:r_h}
    r_h = \frac{d-2}{4\pi} \beta \left(1-\frac{d-2}{8\pi^2} \beta^2\Omega^2+O(\Omega^4)\right).
\end{equation}
To describe the shape of the black hole, we would like to define the black hole eccentricity $e = \sqrt{1-r_{\perp}^2/r_{||}^2}$. There are several ways to define the radii of a rotating black hole~\cite{Ceplak:2024dxm}. We make the following choice, following~\cite{Ceplak:2024dxm}: for the polar radius, we can measure the area of the horizon along the remaining $d-3$ directions and extract the radius, while for the equator radius, we measure the horizon area in the equatorial plane. This choice gives the radii
\begin{equation}
    \begin{split}
        & r_{||} = \sqrt{r_h^2 + a_J^2} = \frac{d-2}{4\pi} \beta \left(1+\frac{d^2-8d+12}{32\pi^2}\beta^2 \Omega^2 + \mathcal{O}(\Omega^4)\right), \\
        & r_{\perp} = r_h = \frac{d-2}{4\pi} \beta \left(1 - \frac{d-2}{8\pi^2} \beta^2 \Omega^2 + \mathcal{O}(\Omega^4) \right),
    \end{split}
\end{equation}
as well as the eccentricity
\begin{equation}
\begin{split}
    e = \frac{2\pi J}{\sqrt{S^2 + 4\pi^2 J^2}}= \frac{d-2}{4\pi} \beta \Omega + O(\Omega^3).
\end{split}
\end{equation}

All the black hole calculations we presented were done classically using Einstein's gravity. Therefore, they are reliable only when the black hole is weakly-curved $\beta \gg l_s$. To compare the behavior with the string star, we nevertheless extrapolate them to the string scale $\beta \sim l_s$, which schematically (up to positive $O(1)$ constants) gives
\begin{equation}\label{eq:corr_bh}
\begin{split}
    M \sim \frac{l_s^{d-2}}{G_N} \left(1-\alpha' \Omega^2 + ...\right), &\quad 
    S \sim \frac{l_s^{d-1}}{G_N} \left(1-\alpha' \Omega^2+ ...\right), \quad 
    r_{||}, r_{\perp} \sim l_s \left(1-\alpha' \Omega^2 + ...\right),\\
    &J \sim \frac{l_s^{d-1}}{G_N} l_s \Omega+ ..., \quad 
    e \sim l_s \Omega.
\end{split}
\end{equation}

Using the results of section \ref{subsec:grand_ss}, we can extrapolate the rotating string star properties \eqref{eq:ss_radii}, \eqref{eq:ecc}, \eqref{eq:ss_mass}, \eqref{eq:ss_entropy}, \eqref{eq:ss_J_I} to the intermediate regime $\alpha' m_0^2 \sim 1$ (or $R^2-R_H^2 \sim \alpha'$)~\cite{Chen:2021dsw} which gives 
\begin{equation}
\begin{split}
    M \sim \frac{l_s^{d-2}}{G_N} \left(1+\alpha' \Omega^2 + ...\right),&\quad
    S \sim \frac{l_s^{d-1}}{G_N} \left(1+\alpha' \Omega^2+ ...\right),\quad
    L_{||},L_\perp \sim l_s \left(1+\alpha' \Omega^2 + ...\right),\\
    &J \sim \frac{l_s^{d-1}}{G_N} l_s \Omega+ ... ,\quad
    e \sim l_s \Omega.
\end{split}
\end{equation}
First, note that for both saddles the perturbative parameter at the correspondence region is $l_s \Omega$. At zeroth order, they qualitatively match, as was already shown in the non-rotating case~\cite{Chen:2021dsw}. The angular momentum and eccentricity are the only quantities with a leading correction in $\Omega$. We see that in both cases, the values match up to $O(1)$ factors between the two saddles, which is the best we can hope for in this situation. We stress that the sign also matches, implying that both saddles rotate in the direction of the angular velocity $\Omega$, and become more oblate in the parallel plane due to the centrifugal force.\footnote{Since at this order $J = \mathcal{I} \cdot \Omega$ on both sides, it also means that the moment of inertia qualitatively agrees between the two saddles.}

At subleading order, we note that the mass and the entropy behave differently on both sides. While for the string star both increase with $\Omega$, they decrease for the black hole. As a result, the overall size of the string star grows with $\Omega$, and shrinks for the black hole.
Being a subleading effect, it is not clear if we should worry about this or not.
To see why, imagine redefining the intermediate region as the location where the BH horizon radius is exactly $r_h=l_s$. Using \eqref{eq:r_h}, this happens at $\beta_{corr} = \frac{4l_s}{d-2}\left(1+\frac{2}{d-2}\alpha'\Omega^2 + ...\right)$.
Of course, we can't really trust this computation using Einstein gravity, but it can help to explain the point.  The correction of $\beta_{corr}$ at order $\alpha'\Omega^2$ also corrects the relations \eqref{eq:corr_bh}. It can be shown analytically that this time, the mass and the entropy increase with $\Omega$, just like the string star. At the very least, this exercise shows that the subleading effects are scheme dependent. A more optimistic view would be that once we look at the proper correspondence region, there's no reason the two saddles won't qualitatively match, even beyond the perturbative analysis.

A more demanding test for the correspondence would be beyond the perturbative expansion, for $l_s\Omega \sim O(1)$. The correspondence point (where $r_h \sim l_s$) will now be affected by $\Omega$ at leading order. It would be interesting to compare the black hole and the string star in this regime as well. Unfortunately, it requires an analytic continuation of the string star saddle to large $\Omega$, which is beyond the scope of our perturbative results.

\subsection{Rotating strings}\label{subsec:corr_free_strings}
As the size of the string star increases, it becomes more diluted, and its internal forces decrease. From the other direction, as a string state turns heavier, its internal gravitational forces grow until they can no longer be neglected or treated perturbatively. We therefore expect a transition between the free string phase and the string star saddle. We will follow section \ref{sec:flat_space_therm} and describe the properties of both sides in the microcanonical ensemble.\footnote{The canonical single-string ensemble suffers from multiple subtleties such as negative specific heat~\cite{Mertens:2015ola}.}
To estimate when the gravitational forces of the string become strong, we can compare the gravitational potential to the free string energy level separation 
\begin{equation}
    \frac{G_N M^2}{L^{d-2}} \sim \frac{1}{l_s}.
\end{equation}
Since in the random-walk approximation (valid for large $l_s M$ and small $J$) $L \sim \sqrt{l_s M}$, we find the breakdown scale ($G_N \sim g_s^2 l_s^{d-1}$)
\begin{equation}\label{eq:breakdown_free_string}
    l_s M_\text{breakdown} \sim g_s^{\frac{4}{d-6}}.
\end{equation}

Let us begin with the free string. The free string analysis of section \ref{sec:flat_space_therm} is valid for $M \ll M_\text{breakdown}$. Extrapolating \eqref{eq:entropy_J_M}, \eqref{eq:L_J_M}, \eqref{eq:ecc_J_M} and \eqref{eq:omega_J_M} to $M\sim M_\text{breakdown}$ gives (up to positive $O(1)$ constants)
\begin{equation}\label{eq:free_string_breakdown_therm}
    \begin{split}
        S-\beta_H M \sim - \log g_s + 1-g_s ^{\frac{8}{d-6}}J^2,  &\quad   L_{\perp} \sim l_s g_s^{\frac{2}{d-6}}(1- g_s^{\frac{8}{6-d}}J^2), \quad L_{||} \sim l_s g_s^{\frac{2}{d-6}}(1+ g_s^{\frac{8}{6-d}}J^2),\\
        & e \sim g_s^{-\frac{4}{d-6}} J, \quad \Omega \sim g_s^{-\frac{8}{d-6}} J /l_s,
    \end{split}
\end{equation}
with $\beta_H$ standing for the non-rotating ($\nu=0$) Hagedorn temperature \eqref{eq:hag_nu_zero}.
In section \ref{subsec:micro_ss}, we studied the microcanonical properties of the rotating string star solution, which are valid for $l_s M$ between $g_s^{\frac{4}{d-6}}$ and $g_s^{-2}$.
Extrapolating the mass in \eqref{eq:ss_omega_micro}, \eqref{eq:ss_entropy_micro}, \eqref{eq:ss_sizes_micro} to the breakdown scale \eqref{eq:breakdown_free_string}, we find for the string star
\begin{equation}\label{eq:ss_breakdown_therm}
    \begin{split}
        S-\beta_H M \sim 1- g_s^{\frac{8}{d-6}}J^2 ,  \quad  &
        L_{\perp} \sim l_s g_s^{\frac{2}{d-6}}(1\pm  g_s^{\frac{8}{6-d}}J^2), \quad 
        L_{||} \sim l_s g_s^{\frac{2}{d-6}}(1\pm  g_s^{\frac{8}{6-d}}J^2), \\ 
        &e \sim g_s^{\frac{4}{6-d}} J, \quad \Omega \sim g_s^{-\frac{8}{d-6}} J/l_s,
    \end{split}
\end{equation}
where the $\pm$ is $+$ for $d=3$ and $-$ for $d=5$. Let us compare the free string \eqref{eq:free_string_breakdown_therm} and the string star \eqref{eq:ss_breakdown_therm} extrapolations to the breakdown scale.
Notice that the perturbative parameter on both sides is $g_s^{\frac{4}{d-6}} J$. At leading order, the two sides qualitatively match.\footnote{The missing $\log g_s$ piece on the string side can be attributed to the fact that we don't compute the 1-loop determinant of the string star.} The only quantities that have a leading order piece ($\Omega$ and $e$) have the same $g_s$-dependence and the same sign. 

At subleading order, the corrections to the sizes generally have a different sign. Similarly to the black hole transition region, this doesn't necessarily indicate a problem and depends on the $J$ dependence of the breakdown region. One way to define the breakdown region is by demanding the string star's on-shell action to be $O(1)$ (some $J$-independent constant)~\cite{Chen:2021dsw}. In this way, it is possible to show that the sizes of both the free strings and the string star get a positive correction in $J^2$ at the breakdown scale. With better control over the proper correspondence region, one could hope for a match beyond the perturbative expansions.\footnote{For the regime $(R-R_H(\nu))/l_s \sim |\nu|$ mentioned in section \ref{subsec:eom_ss}, the breakdown of the semiclassical solution happens not at a specific temperature, but for $|\nu| \lesssim g_s^{\frac{4}{6-d}}$. At that point, $M \sim g_s^{\frac{4}{d-6}}$, $\mathcal{I}\sim M^2$, $J \sim i M$, so one might speculate that this phase is connected to free strings with $J\sim M$. This regime is not connected to the black hole phase due to the condition $\beta-\beta_H \lesssim |\nu|\ll 1$.}

A more demanding test for the correspondence would be to analyze the situation at $J \sim g_s^{\frac{4}{d-6}}$, or $J \gg g_s^{\frac{4}{d-6}}$, corresponding to $J\sim M$ or $J\gg M$ in the free string description. The correspondence point should then be affected at leading order in $J$. Indeed, for $J\sim M^2$, we are in the ultraspinning regime, which has been analyzed in~\cite{Ceplak:2023afb,Ceplak:2024dxm}. Again, this requires more control over the string star solution than we currently have.

\subsection{Large angular momentum and extremal objects}
The rotating string star solution was analyzed in section \ref{sec:sss} for perturbative angular momenta in the near-Hagedorn temperature regime.
In terms of $M$ and $J$, we only considered the regime where $J/(l_s M) \ll 1$ for large $M$. However, notice this was merely a simplifying assumption. The EFT is valid in a larger domain \eqref{eq:eft_valid}, which translates to $J/(l_s M) \sim 1$.
Although beyond the scope of our calculations, it is interesting to speculate what happens for higher angular momenta $J \sim l_s M \gg 1$.
 
From \eqref{eq:ss_entropy_micro}, we still have the entropy-energy relation $S \approx \beta_H M$ at leading order in $M$. At subleading order \eqref{eq:ss_entropy_micro}, we expect the entropy to shrink. Since the self-gravitating strings still carry Hagedorn entropy, the eccentricity of the solution will grow, but remain finite (far enough from unity).
This behavior is also expected by comparison to both free strings and black holes. Looking at \eqref{eq:free_string_breakdown_therm}, we expect a Hagedorn growth to remain at $J/M \sim 1$. Furthermore, in~\cite{Ceplak:2024dxm}, the size and eccentricity of free strings for various values of $J/M$ were computed, finding finite eccentricity. Since rotating black holes have finite eccentricity for any $M$,$\Omega$, we could expect a correspondence principle in this regime as well.

It is tempting to speculate about the string star solution closer to the string Regge bound $J \lesssim \frac{1}{2} \alpha' M^2$ \eqref{eq:J_bound}~\cite{Ceplak:2023afb,Ceplak:2024dxm}. 
Lacking control over such a large $\Omega$ regime, it is hard to predict the result.
A natural suggestion would be that in this regime, the string star is composed of near-extremal strings, with small $O(1)$ entropy. In this picture, one could expect the eccentricity to be very close to unity, with $L_\perp \sim l_s$. This `stringy pancake' limit is likely to transition into a `quantum string bar'. It would be interesting to better understand this regime.

Finally, one could wonder about the relation between our solution to extremal rotating black holes, which also satisfy $J \propto M^2$ (albeit with a different $g_s$ factor~\cite{Ceplak:2023afb}). The extremal limit necessarily involves arbitrarily low temperatures and a very large horizon area. As a result, we believe that in order to potentially deform an (Eudlidean) extremal black hole to a string star, one has to first go through higher temperatures $\beta \sim l_s$.

\section{Anti-de Sitter space}
\label{sec:ads}
In this section, we generalize the analysis from flat space to anti-de Sitter space. Consider an AdS$_{d+1}\times M_{9-d}$ string theory background for some compact $M_{9-d}$, and its holographic CFT$_d$ on $S^{d-1}\times \mathbb{R}$. Unless said otherwise, we will assume throughout the section that the background is type II with no NS-NS fluxes supported in the time directions. We believe that generalizing our results beyond those assumptions can be straightforward~\cite{Urbach:2023npi}.
In terms of the holographic CFT, consider the partition function
\begin{equation}\label{eq:Z_CFT_def}
    Z(\beta,\Omega) = \text{Tr}_\text{CFT}
    \left(\exp(-\beta \left(H - \Omega J\right))
    \right),
\end{equation}
with $J$ being a rotation generator in $S^{d-1}$.\footnote{Notice that here both $\beta$ and $\Omega$ are defined on the CFT side in dimensionless units. In terms of the dual string theory, they are in $l_{ads}$ units. As a result, they are related to the previous section's $\beta$, $\Omega$ by factors of $l_{ads}$.}
This ensemble is (holographically dual to) the AdS analog of \eqref{eq:beta_nu_ens}. Unlike in flat space, where one fixes the asymptotic length of the thermal circle, in AdS one fixes the asymptotic form of the metric (or equivalently the conformal boundary)\cite{Hawking:1982dh,Witten:1998qj}. This prescription renders the AdS gravitational partition function well defined even at finite string coupling, in contrast to the situation in flat space.

Similar to flat space, we focus our attention on two saddles that contribute to \eqref{eq:Z_CFT_def}: the rotating thermal AdS saddle, which is the closest analog of the thermal $\mathbb R^d \times S^1$ in flat space, and the Euclidean rotating AdS black hole. In what follows, we first discuss the Hagedorn temperature in rotating thermal AdS and then comment on the AdS rotating string star saddle.

\subsection{The Hagedorn temperature of rotating thermal AdS}
The gravity dual of \eqref{eq:Z_CFT_def} includes geometries with a twisted $S^{d-1} \times S^1_{\beta}$ as conformal boundary.
Below the Hawking-Page transition, the dominant bulk geometry is the rotating thermal AdS geometry
\begin{equation}
    ds^2 = R^2 \cosh^2\left(r/l_{ads}\right) d\tau^2 + dr^2 + l_{ads}^2 \sinh^2\left(r/l_{ads}\right) d\Omega_{d-1}^2, 
\end{equation}
with $R = \beta/(2\pi) \cdot l_{ads}$, the bulk thermal radius. Following the flat-space analysis, we write the sphere's metric as $d\Omega_{d-1}^2 = d\theta^2 + \sin^2 \theta \ d\phi^2 + \cos^2\theta \ d\Omega_{d-3}^2$, with $\phi$ the angle generated by $J$. In these coordinates, the boundary conditions identify $(\tau,\phi) \sim (\tau+2\pi,\phi + i \beta \Omega) \sim (\tau,\phi+2\pi)$. In the limit $g_s=0$ the thermal AdS solution remains dominant to arbitrarily high energies. On the dual CFT side, this is the $N=\infty$ limit of the theory (the confined phase). In this limit, we would like to determine the temperature $R_H(\Omega)$ at which the partition function \eqref{eq:Z_CFT_def} diverges.

As was shown in~\cite{maldacena_private,Urbach:2022xzw}, the leading correction in $l_s/l_{ads}$ of the Hagedorn temperature can be computed using the effective action \eqref{eq:eft_chi2}, adapted to curved space
\begin{equation}\label{eq:eft_chi_ads}
    S = \int d^D x \sqrt{G} \left(|\partial \chi|^2 + m^2 |\chi|^2 \right),
\end{equation}
with 
\begin{equation}\label{eq:chi_mass_ads}
\begin{split}
    m^2 &= -\frac{2}{\alpha'}+\frac{1}{(\alpha')^2}\left(R^2 \cosh^2(r/l_{ads})-R^2 \Omega^2 \sinh^2(r/l_{ads}) \sin^2\theta\right)\\
    &=-\frac{2}{\alpha'}+\frac{1}{(\alpha')^2}\left(R^2 + \frac{R^2}{l_{ads}^2}(r^2-\Omega^2 r^2\sin^2\theta) + ...\right),
\end{split}
\end{equation}
where in the second line we expanded to quadratic order in $r$.
We already see a dramatic difference compared to flat space. Due to the AdS curvature, the winding mode remains stable even with a real chemical potential, as long as $\Omega<1$. The instability of the ensemble for $\Omega\ge 1$ is a direct result of the unitarity bound $E \ge J$ (in AdS units)~\cite{Hawking:1998kw,Hawking:1999dp,Witten:2021nzp,Kim:2023sig}.

The Hagedorn temperature is the temperature at which the lowest eigenvalue of the quadratic action becomes a zero mode, or satisfies the equations of motion.
The leading order in $l_s/l_{ads}$ (for fixed $\Omega$) is given by taking the flat space kinetic term, but expanding the mass to leading order, giving the Schrodinger-like equation
\begin{equation}\label{eq:eom_chi_ads}
    -\nabla^2 \chi +\frac{R^2}{(\alpha')^2 l_{ads}^2}\left(r^2-\Omega^2 r^2 \sin^2\theta\right)\chi  = \left(\frac{2}{\alpha'}-\frac{R^2}{(\alpha')^2} \right)\chi.
\end{equation}
It is useful to move to cylindrical coordinates $\rho = r \sin \theta$ and $z=r \cos \theta$. In these coordinates, the equation factorizes, which allows us to solve for the lowest normalizable eigenvalue
\begin{equation}
    \chi(r,\theta) = \exp\left(-\frac{R}{2\alpha' l_{ads}}\left(r^2 \cos^2\theta+\sqrt{1-\Omega^2} r^2 \sin^2\theta \right)\right).
\end{equation}
Setting to leading order $R \sim l_s$, the characteristic scales of the solutions are
\begin{equation}
    L^2_{||} \sim \frac{l_s l_{ads}}{\sqrt{1-\Omega^2}}, \quad L^2_{\perp} \sim l_s l_{ads}.
\end{equation}
Following the Newtonian intuition, as we increase the angular velocity $\Omega$, the solution becomes wider in the 2d plane. When $1-\Omega^2 \sim l_s/l_{ads}$, we expect higher order kinetic terms to be leading in the plane, eventually leading to an $L_{||} \sim l_{ads}$ sized solution~\cite{Urbach:2023npi}.

Comparing with the RHS of \eqref{eq:eom_chi_ads} gives an equation for the AdS Hagedorn temperature
\begin{equation}\label{eq:ads_hag_eq}
    R_H^2 /\alpha' = 2 -\frac{R_H}{l_{ads}} \left(d-2+2\sqrt{1-\Omega^2}\right).
\end{equation}
When are we supposed to trust \eqref{eq:ads_hag_eq}? From the CFT perspective, it is natural to keep $\Omega$ fixed, but expand the answer in orders of $l_s/l_{ads}$ (corresponding in the CFT to a strong coupling expansion). Understood in this way, we can trust our calculations only up to order $l_s/l_{ads}$. Expanding \eqref{eq:ads_hag_eq} to this order leads to
\begin{equation}\label{eq:hag_ads_omega}
    R_H /l_s = \sqrt{2} -\frac{l_s}{2l_{ads}} \left(d-2+2\sqrt{1-\Omega^2}\right)+ O(\alpha'/l_{ads}^2).
\end{equation}

A second limit we can take is for imaginary chemical potential $\beta \Omega = i 2\pi \nu$, or $\Omega = i l_{ads}/R \cdot \nu$. We can imagine a fixed $\nu$ just like in the flat space case, and expand in $l_s/l_{ads}$. 
Expanding \eqref{eq:ads_hag_eq} for fixed $\nu$ up to order $l_s/l_{ads}$ gives
\begin{equation}\label{eq:hag_ads_nu}
    R_H/l_s = \sqrt{2-2|\nu|} - (d-2)\frac{l_s}{2l_{ads}} + O(\alpha'/l_{ads}^2).
\end{equation}
The zeroth order $R_H/l_s=\sqrt{2-2|\nu|}$ agrees with the flat space computation above \eqref{eq:hag_flat_final}.
This expansion is self-consistent as long as $l_s/l_{ads} \ll |\nu|<1$. In this case, the sizes of the wavefunction are
\begin{equation}
    L^2_{||} \sim \frac{\alpha'}{|\nu|}, \quad L^2_{\perp} \sim l_s l_{ads}.
\end{equation}
As we can see, the solution size in the 2d plane is significantly smaller compared to the perpendicular direction $L_{||} \ll L_{\perp}$. The intuition is that in the 2d plane, the solution is effectively the flat space solution, giving the first term in \eqref{eq:hag_ads_nu}. In the other $d-2$ directions, the solution is the standard AdS result \cite{maldacena_private,Urbach:2022xzw}, which leads to the second term in \eqref{eq:hag_ads_nu}. It would be interesting to check how higher-order corrections depend on $\nu$. 

Finally, we could consider a different limit where $\nu \ll l_s/l_{ads}$. In this limit, $\nu$ is a perturbation of the full AdS computation. Since we don't have control over the $\nu=0$ behavior for finite $l_s/l_{ads}$, it is difficult to analyze the leading correction to it by $\nu$. We therefore cannot say anything valuable about this limit.

\paragraph{Comments on AdS\titlemath{$_3$}}
In AdS$_3$ ($d=2$), we can consider backgrounds with non-zero NS-NS flux $\lambda$~\cite{Cho:2018nfn,Urbach:2023npi}. To leading $\alpha'$ order, the effect of $\lambda$ on the quadratic action of the winding mode is the same as $\Omega$. Shifting $\Omega^2 \mapsto \Omega^2+\lambda^2$ in \eqref{eq:hag_ads_omega} gives the leading correction to the Hagedorn temperature
\begin{equation}
    R_H/l_s = \sqrt{2} - \frac{l_s}{l_{AdS}}\sqrt{1-\Omega^2-\lambda^2} + \mathcal{O}\left(\frac{\alpha'}{l_{AdS}^2}\right).
\end{equation}
Notice that this time, the ensemble is unstable for $\Omega^2+\lambda^2 \ge 1$ due to condensation of rotating long-strings at large radii.
Following~\cite{Harmark:2024ioq,Canneti:2025cos}, it would be interesting to study higher order corrections to the Hagedorn temperature at order $1/l_{AdS}^2$, $1/l_{AdS}^3$ as a function of the NS-NS flux $\lambda$ and angular momentum $\Omega$. In~\cite{Harmark:2024ioq}, the higher $\alpha'$-corrections to the winding mode equation were found by reverse-engineering from the exactly known Hagedorn temperature on the pp-wave background. As they are stated in covariant form, one can also use these expressions to find the $\alpha'$-corrections to the Hagedorn temperature in the ensemble with nonzero $\nu$. 

The analysis above breaks for the pure NS-NS background $\lambda=1$. At $\Omega=0$, the magnetic flux and the gravitational contributions to the effective mass cancel. As a result, the winding mode has no potential and extends with $L \sim l_{ads}$~\cite{Urbach:2023npi}. Fortunately, the pure NS-NS background has an RNS description, which allows for a direct calculation of the spectrum (see the recent~\cite{Ferko:2024uxi}).
The string spectrum of bosonic strings in thermal pure NS-NS AdS$_3$ for general $\beta, \nu$ was studied in~\cite{Mertens:2014nca}, where the authors also found the Hagedorn temperature $R_H(\nu)$. Modifying their computation for type II strings, we find ($-1<\nu<1$)
\begin{equation}\label{eq:R_H_ads3_pure_nsns}
    R_H^2/\alpha' = 
    \begin{cases}
        2-\frac{1}{k}-k \nu^2 & |\nu| < 1/k\\
        2-2|\nu| & 1/k <|\nu| <1
    \end{cases},
\end{equation}
with $k=l_{ads}^2/\alpha'$.
Surprisingly, this temperature agrees exactly with the flat space result \eqref{eq:ehag_II} for $k |\nu| >1$. Thus, it also agrees for any $\nu$ in the flat space limit $k \gg 1$. It also converges to the known result at $\nu=0$~\cite{Berkooz:2007fe}. The (second-derivative) discontinuity of $R^2_H$ at $k |\nu| = 1$ is the location where the $SL(2,R)$ representation of the winding mode turns from continuous at $k |\nu| < 1$ to discrete at $k |\nu| >1$. As we increase $\nu$, the size of the winding mode gets shorter. We interpret it as the transition where generic strings turn from long strings to short strings~\cite{Maldacena:2000hw}. It would be interesting to derive \eqref{eq:R_H_ads3_pure_nsns} directly 
using the recent results of~\cite{Ferko:2024uxi}.

\subsection{Rotating AdS string stars}
In~\cite{Urbach:2022xzw}, a string star saddle was found numerically in AdS space. The solution size was shown to grow with the temperature, with a maximal size of $L^2 \sim l_s l_{ads}$ at the AdS Hagedorn temperature. At lower temperatures $l_s/l_{ads} \ll (R-R_H)/R_H \ll 1$ (and $3\le d\le 5$), the star's size is much smaller than the AdS radius, and thus approximately given by the flat space string star configuration. Similarly, the (small) AdS black hole at high temperatures $l_s \ll \beta \ll l_{ads}$ is well approximated by the Schwarzschild solution. As a result, the qualitative arguments for the string/black hole correspondence at intermediate temperatures $R-R_H \sim l_s$ in flat space immediately apply to AdS.

It is natural to look for a rotating string star solution in AdS as well. This will require expanding the string star action \eqref{eq:eft_chi_ads} to cubic order in the fields around the $\nu$ (or $\Omega$) dependent thermal AdS geometry. Although we will not do it explicitly here, it is simple enough to estimate the behavior of the solution.

Expanding in small $\nu \ll 1$, we expect a perturbation of the non-rotating AdS string star. 
To leading non-trivial order, we can use the rigid-body approximation \eqref{eq:ss_J_I_2}. 
Using~\cite{Urbach:2022xzw}, we can estimate the moment of inertia close to the AdS Hagedorn temperature $(R-R_H)/R_H \ll l_s/l_{ads}\ll 1$:
\begin{equation}
    \mathcal{I} \sim \frac{(l_s l_{ads})^{d/2}}{G_N} \frac{R-R_H}{R_H}.
\end{equation}
Near the AdS Hagedorn temperature, we also expect the solution to have a size similar to that of the winding mode described in the previous section.
Note that due to the AdS curvature attraction, it is possible in this case to find real $\Omega$ solutions for the string star directly, without the need to go to imaginary angular velocity.
At lower temperatures, however, we expect the size to shrink. For $l_s/l_{ads} \ll (R-R_H)/R_H \ll 1$ and $3\le d\le 5$ (and real $\nu$), the solution coincides with the flat-space rotating string star solution of section \ref{sec:sss}. As in $\nu=0$, we thus expect the arguments of section \ref{sec:corr} to follow to AdS as well.\footnote{For recent analyses of the superradiant instability of rotating AdS black holes and the proposed end-state in the form of ‘grey galaxies’ (mixed configurations of a near-critical $\Omega \sim 1$ black hole and a co-rotating thermal cloud) see~\cite{Kim:2023sig,Bajaj:2024utv}.}

A nuanced version of this discussion exists for AdS$_3$. While in $d=2$ flat space no string star solutions exist, it was shown in~\cite{Urbach:2023npi} that such solutions do exist in AdS$_3$. Following~\cite{Halder:2022ykw}, it was suggested that the AdS$_3$ string star is continuously related to the `S-transformed' BTZ solution deformed by an angular winding condensate (defined only for very low temperatures). Adding non-zero angular velocity, we expect a generalization of this discussion. In terms of the modular parameter $\tau = \nu + i \beta/(2\pi)$, we can define the Hagedorn line $\tau_H = \nu + i R_H(\nu)/l_{ads}$. A reliable string star solution exists in a small region close to $\tau_H$.\footnote{More precisely, $\tau= \nu + i R/l_{ads}$ is in the region if $\nu \ll 1$ and $R-R_H(\nu) \ll l_s$.} Using the S-transformation $\tau'=-1/\tau$, there also is a second region close to $-1/\tau_H$ where we can find a reliable winding condensate, this time around the (rotating) BTZ solution. The string/BTZ transition suggested in ~\cite{Halder:2022ykw,Urbach:2023npi} can be understood as the claim that these two regions are smoothly related as a function of $\tau$.

We can also consider a more general transformation than just S. Of the entire modular group $SL(2,\mathbb{Z})$, only transformations that keep the boundary condition of the fermions map a string star solution to another one.\footnote{Changing the target-space fermion boundary condition amounts to a different GSO projection, and hence a different winding spectrum. This is the same as shifting $\nu$ to $\nu'=\nu+1$.}
The subgroup that fixes the boundary conditions is generated by $S$ and $T^2$, and goes by the name the theta group $\Gamma_\theta$~\cite{schoeneberg2012elliptic}. For its fundamental domain, we can take $|\tau|>1$ with $-1 < \Re \tau <1$. In this domain, the reliable winding condensate is an angular condensate of a very cold ($|\tau| \sim l_{ads}/l_s$) rotating BTZ solution.
A natural generalization of the string/BTZ transition suggestion would be that it is possible to extend the winding condensate solution in such a way that $\Gamma_\theta$ maps the solutions at the boundaries of the fundamental region to each other. If true, it means that all the $\Gamma_\theta$ mappings of the string star solution are continuously related to each other, in the same way the AdS$_3$ string star is related to the BTZ solution.

\section{Future directions}
There are several natural generalizations of our work. One is to include more than one angular momentum (for $d>3$), and compare with the appropriate (Myers–Perry) black hole. Specifically for $d=4$, one can turn on equal angular momenta for the $1,2$ and the $3,4$ directions, in which the string star equations are one-dimensional again (as in $d=2$). It would also be interesting to better understand the AdS rotating solutions, especially for pure NS-NS, where one could hope for an exact worldsheet description following~\cite{Agia:2023skp}.
In AdS, it is also common to turn on chemical potentials for rotations of internal direction~\cite{Chamblin:1999tk} (dual on the CFT to global symmetries). Looking for an AdS string star rotating in those directions will generalize the charged solutions of~\cite{Chen:2021dsw}.

In~\cite{Harmark:2021qma}, an integrability analysis of $\mathcal{N}=4$ SYM allowed for the computation of the CFT Hagedorn temperature even at strong coupling, with an exact matching with the bulk $l_s/l_{ads}$ expansion. Using the same methods, the authors studied the Hagedorn temperature in the presence of chemical potentials for angular momentum of the type \eqref{eq:Z_CFT_def}. It would be illuminating to compare the strong coupling regime with \eqref{eq:hag_ads_omega}. Following the recent progress~\cite{Bigazzi:2023oqm,Ekhammar:2023glu,Bigazzi:2023hxt,Ekhammar:2023cuj,Harmark:2024ioq,Bigazzi:2024biz,Bigazzi:2024sjy,Canneti:2025rsp,Castellano:2025ljk} one could study higher $l_s/l_{ads}$ orders of the Hagedorn temperature, either $R_H(\Omega)$ \eqref{eq:hag_ads_omega} or $R_H(\nu)$ \eqref{eq:ads_hag_eq}.
Following the analysis of section \ref{sec:ads}, it is also possible to study the Hagedorn temperature and rotating winding strings condensates in the presence of an (holographic) angular twist $\nu$ in confining backgrounds~\cite{Urbach:2023npi}.

More conceptually, we hope for a better understanding of the rotating string star solution at large angular momentum, and its possible transition to the quantum string rod~\cite{Ceplak:2023afb}. One could also wonder if there is a string star version for the extremal black hole, in the double limit of zero temperature and $J \sim g_s^{-\frac{4}{d-1}}$ (at which $r_h \sim l_s$).

\section*{Acknowledgments}
We would like to thank
Ofer Aharony,
Yiming Chen,
Roberto Emparan,
Juan Maldacena,
Ohad Mamroud,
Andrea Puhm, and
Marija Toma\^sevi\'c
for useful discussions.
We especially thank Suman Kundu for his involvement and collaboration in the initial stages of this work. 
JS would like to thank the University of Boulder Colorado and the Institute for Advanced Study for their hospitality while part of this work was carried out. EYU also thanks the University of Amsterdam for its hospitality. 
The work of JS is supported by a research grant from the Chaim Mida Prize in Theoretical Physics at the Weizmann Institute of Science, and by a Dean Award of Excellence of the Weizmann School of Science. The work of JS was also supported in part by ISF
grant no. 2159/22, by Simons Foundation grant 994296 (Simons Collaboration on Confinement and QCD Strings), by the Minerva foundation with funding from the Federal German
Ministry for Education and Research, and by the German Research Foundation through a
German-Israeli Project Cooperation (DIP) grant “Holography and the Swampland”. EYU is supported by the J. Robert Oppenheimer Endowed Fund.

\appendix

\section{Worldsheet state counting}
\label{app:state_count}
In this appendix, we derive properties of highly-excited rotating strings by directly studying the full worldsheet partition function. We will work in light-cone quantization and use $\alpha'=1$. 
In section \ref{app:hagedorn_temps}, we derive the $\nu$-dependent Hagedorn temperature for bosonic, type II, and heterotic strings. In section \ref{app:micro_entropy}, we derive equation \eqref{eq:S_M_J} of the microcanonical entropy in the $J/M^2$ scaling. Lastly, in section \ref{app:J/M free string} we derive properties of highly excited free strings in the $J/M \ll 1$ expansion.

\subsection{Hagedorn temperatures}\label{app:hagedorn_temps}
In this section, we begin by computing the worldsheet partition function ($\tau=\tau_1+i \tau_2$)
\begin{equation}
    Z(\tau,\nu) = \text{Tr}_\text{WS}\left(\exp\left(2\pi i \left(\tau_1 P + \tau_2 H + \nu J\right)\right) \right),
\end{equation}
of different string theories at a given chirality. Setting $\tau=i b/(2\pi)$\footnote{$b$ is the worldsheet temperature, which is distinct from the spacetime temperature $\beta$.}, we will be interested in the exponential growth in the high-energy limit $b \ll 1$ at fixed $\nu$
\begin{equation}\label{eq:ass_growth}
    Z(\tau=i b/(2\pi),\nu) \sim \exp\left(\frac{2\pi^2}{b} a(\nu) + O(b^0)\right).
\end{equation}
As we showed in section \ref{subsec:grand_can}, standard thermodynamics arguments (and the mass-shell relation $H \approx M^2/2$) show that the coefficient $a(\nu)$ is related to the spacetime Hagedorn radius by $R_H^2(\nu) = a(\nu)$. 
Since the worldsheet theory is essentially free in each chirality, each worldsheet field contributes additively to $a(\nu)$. Given $a_R$ and $a_L$ from each chirality, the Hagedorn growth from both satisfies $R_H = \frac{1}{2} \left(a_R^{1/2}+a_L^{1/2}\right)$.

At $\nu=0$, the behavior \eqref{eq:ass_growth} is determined by the Cardy formula, which relates the worldsheet central charge to the spacetime Hagedorn growth. For the same reason, fields that carry zero angular momentum ($J=0$) contribute according to the Cardy formula
\begin{equation}
    a_0 = \frac{c}{6},
\end{equation}
with $c$ being the corresponding central charge. For a single scalar ($c=1$) or spinor ($c=1/2$)
\begin{equation}
    a_{B,0}(\nu) = \frac{1}{6}, \quad a_{F,0}(\nu) = \frac{1}{12}.
\end{equation}

We now turn to the spin $1$ fields, beginning with the complex boson $\Phi = X^1 + i X^2$.
Diagonalizing $H, P,$ and $ J$ together gives the eigenvalues $J=\pm1$ for each oscillator~\cite{Russo:1994ev,Matsuo:2009sx}. The complex boson partition function of a single chirality gives
\begin{equation}
\begin{split}
    Z_{B,1}(\tau,\nu) &= \prod_{n=1}^\infty \frac{1}{
    \left(1-e^{2\pi i (\tau n+ \nu)}\right)
    \left(1-e^{2\pi i (\tau n- \nu)}\right)
    }\\
    &= \frac{-2 e^{i\frac{\pi}{6}\tau} \eta(\tau) \sin(\pi\nu)}{ \vartheta_{11}(\nu,\tau)}\\
    &= \frac{2 i e^{i\frac{\pi}{6}\tau + i \pi \nu^2/\tau} \eta(-1/\tau) \sin(\pi\nu)}{ \vartheta_{11}(\nu/\tau,-1/\tau)},
\end{split}
\end{equation}
where in the second line we use the modular properties of $\vartheta_{11},\eta$~\cite{Polchinski:1998rq}.
To find the high-temperature behavior, we set $b = -2\pi i \tau$ and focus on the exponential form in the $b \ll 1$ limit. For brevity, we will assume the fundamental region $-1/2<\nu<1/2$.
The $\eta$ and $\vartheta_{11}$ functions satisfy in the $e^{2\pi i \cdot -1/\tau} = e^{-4\pi^2/b}\rightarrow 0$ limit the identities
\begin{equation}\label{eq:ass_theta}
\begin{split}
    \eta(-1/\tau) &\sim \exp\left(-\frac{\pi^2}{6 b}\right),\\
    \vartheta_{11}(\nu/\tau,-1/\tau) &\sim 2i\sinh(2\pi^2 \nu/b) e^{-\frac{\pi^2}{2b}},
\end{split}
\end{equation}
which together give the asymptotic behavior
\begin{equation}
\begin{split}
    Z_{B,1}(b,\nu) &\sim e^{-\frac{b}{12} +\frac{\pi^2}{3b}+ \frac{2\pi^2}{b}\nu^2} \frac{\sin(\pi\nu)}{\sinh(2\pi^2 \nu/b)}
    .
\end{split}
\end{equation}
The asymptotic growth is therefore (summing up the two chiralities)
\begin{equation}
    a_{B,1}(\nu) = \frac{1}{3}- 2(|\nu|-\nu^2).
\end{equation}

The bosonic string theory spectrum includes $22$ trivial bosons and a single vector boson, which together give
\begin{equation}\label{eq:hagedorn_bosonic}
\begin{split}
    R^2_H(\nu) &= 22 a_{B,0} + a_{B,1}\\
    &= 4 - 2(|\nu|-\nu^2).
\end{split}
\end{equation}

Moving to type II string theory, the new component is, of course, from the GSO-projected worldsheet fermions. The GSO projection is best understood in the case of periodic boundary conditions for the target space fermions. While
$\nu=0$ stands for the anti-periodic boundary conditions, the periodic case is exactly $\nu=1$. We therefore write down the standard GSO-projected partition function~\cite{Polchinski:1998rr}, only with angular velocity $\hat \nu = 1-\nu$ ($-1<\nu<1$) (see~\cite{Ferko:2024uxi} for a recent discussion)
\begin{equation}
\begin{split}
    Z_{\psi}(\nu) &= \frac{1}{2}\left(
     Z_0^0(\tau)^3 Z_0^0(\tau,\hat \nu)
    -Z_1^0(\tau)^3 Z_1^0(\tau,\hat \nu)
    -Z_0^1(\tau)^3 Z_0^1(\tau,\hat \nu)
    +Z_1^1(\tau)^3 Z_1^1(\tau,\hat \nu)
    \right).
\end{split}
\end{equation}
Here $Z^\alpha_\beta(\tau)=Z^\alpha_\beta(\tau,\nu)$ (not to be confused with the target space temperature $\beta$), and 
\begin{equation}
\begin{split}
    Z_\alpha^\beta(\tau,\nu) &= \text{Tr}_\alpha \left(e^{i \pi \beta F} e^{2\pi i (\tau H+\nu J)}\right)\\
    &= \prod_{n=1}^\infty 
    \left(1+e^{i\pi \beta} e^{2\pi i (\tau (n-\frac{1-\alpha}{2})+ \nu)}\right)
    \left(1+e^{-i\pi \beta}e^{2\pi i (\tau (n-\frac{1-\alpha}{2})- \nu)}\right)
    \\
    &= \frac{\vartheta_{\alpha \beta}(\hat \nu,\tau)}{e^{-i \frac{\pi}{12}\tau} \eta(\tau)}.
\end{split}
\end{equation}
This leads to
\begin{equation}
\begin{split}
    Z_\psi (\nu) &= \frac{e^{\frac{\pi}{3}\tau}}{2\eta(\tau)^{4}}\left(
    \theta_{00}(\tau)^3\theta_{00}(\hat \nu,\tau)
    -\theta_{01}(\tau)^3\theta_{01}(\hat \nu,\tau)
    -\theta_{10}(\tau)^3\theta_{10}(\hat \nu,\tau)
    +\theta_{11}(\tau)^3\theta_{11}(\hat \nu,\tau)\right)\\
    &= 
    \frac{e^{\frac{\pi}{3}\tau}}{\eta(\tau)^{4}}
    \theta_{11}(\hat \nu/2,\tau)^4.
\end{split}
\end{equation}
Using the modular properties of $\theta_{11}$,
\begin{equation}
    Z_\psi(\nu) = e^{\frac{\pi}{3}\tau-\frac{i\pi \hat \nu^2}{\tau}}\frac{\theta_{11}(\hat\nu/(2\tau),-1/\tau)^4}{\eta(-1/\tau)^4},
\end{equation}
and \eqref{eq:ass_theta}, setting $\tau=i b/(2\pi)$ we find that $Z_\psi$ has the high-energy behavior ($-1<\nu<1$)
\begin{equation}
    a_{F,II} = \frac{2}{3}-2\nu^2.
\end{equation}
As a result, the type II Hagedorn temperature is
\begin{equation}
    R_H^2/\alpha' = 6 a_{B,0} + a_{B,1} + a_{F,II} = 2-2|\nu|.
\end{equation}

The heterotic case combines both computations.
The right movers include the same content as type II, with the same answer
\begin{equation}
\begin{split}
    a_R &= 2-2|\nu|.
\end{split}
\end{equation}
The left-movers content includes $6$ trivial bosons, $32$ trivial spinors, and $1$ vector boson, leading to
\begin{equation}
\begin{split}
    a_L &= 6 a_{B,0} + 32 a_{F,0} + a_{B,1} \\
    &= 4-2(|\nu|-\nu^2).
\end{split}
\end{equation}
Taking the sum of the two linear growths leads to the heterotic Hagedorn radius
\begin{equation}\label{eq:hagedorn_het}
\begin{split}
    R_{H} &= \frac{1}{2} \left(a_R^{1/2} + a_L^{1/2}\right)\\
    & = \frac{1}{2} \left(\sqrt{2-2|\nu|}
    +
    \sqrt{4-2|\nu|+2\nu^2)}
    \right).
\end{split}
\end{equation}

\subsection{Entropy at fixed \titlemath{$J/M^2$}}
\label{app:micro_entropy}
In this section, we would like to study the microcanonical entropy $S(M,J)$ in the scaling $M^2 \gg 1$ for fixed $J/M^2$, mostly following \cite{Matsuo:2009sx}.
To do that, we begin by writing down the worldsheet partition function in the high temperature limit $b\ll 1$ for a given chirality. For any of the string theories we described, it has the form 
\begin{equation}\label{nu b partition function}
    Z(b,\nu) \sim \exp\left(\frac{\beta_H^2}{4b} (1 + O(\nu^2)) + O(b^0)\right) \frac{\sin(\pi\nu)}{\sinh(2\pi^2 \nu/b)} 
\end{equation}
The inverse Fourier transform in $\nu$ will be dominated by the $\nu \sim b \ll 1 $ limit, in which the $O(\nu^2)$ terms are subleading. In this limit, we find
\begin{equation}
\begin{split}
    Z(b,j) &= \int_{-1/2}^{1/2} d\nu e^{2\pi i \nu j} Z(b,\nu)\\
    &= e^{\frac{\beta_H^2}{4b}(1 + O(\nu^2)) + O(b^0)} \int_{-1/2}^{1/2} d\nu e^{2\pi i \nu j} \frac{\sin(\pi\nu)}{\sinh(2\pi^2 \nu/b)}.
\end{split}
\end{equation}
The integral is concentrated around $\nu \sim b$, where the non-universal terms vanish. To see that, we define $x = \nu/b$. We can write the integral in the new variable,
\begin{equation}
\begin{split}
    Z(b,j) &= b\cdot \int_{-\frac{1}{2b}}^{\frac{1}{2b}} dx e^{\frac{\beta_H^2}{4b} + O(b x^2) + O(b^0))} e^{2\pi i b j x} \frac{\sin(\pi b x)}{\sinh(2\pi^2 x)}\\
    & = e^{\frac{\beta_H^2}{4b}} \pi b^2\cdot \int_{-\infty}^{\infty} dx e^{2\pi i (b\cdot j) \cdot x} \frac{x}{\sinh(2\pi^2 x)}\\
    & = e^{\frac{\beta_H^2}{4b}} \frac{b^2}{8\pi \cosh^2(b j/2)}.
\end{split}
\end{equation}
Let us further concentrate on the regime where $j \sim 1/b^2$, in which we have at leading $b \ll 1$ order
\begin{equation}
    \log Z(b,j) = \frac{\beta_H^2}{4b}-b |j| + O(b^0).
\end{equation}
Standard thermodynamics gives $\langle N\rangle = \beta_H^2/(4b^2) + |j| + O(b^0)$ (notice that $j$ is the angular momentum from a given chirality). To find the entropy from a given chirality, we also use the mass shell relation $N \sim \frac{1}{4}M^2$ to find
\begin{equation}
    s(M,j) = \frac{\beta_H^2}{2b} + O(b^0, (b^2 j)^0)  = \frac{1}{2}\beta_H \sqrt{M^2 - 4|j|} + O(M^0).
\end{equation}
The mass shell condition also shows that the regime we choose, $j \sim 1/b^2$, is indeed equivalent to fixing $j / M^2$ at large $M$.

To find the total entropy from both chiralities, we need to extremize over
\begin{equation}
    S(M,J) = \max_{j_R+j_L=J} \left(\frac{1}{2}\beta_{H,L} \sqrt{M^2 - 4|j_L|} + \frac{1}{2}\beta_{H,R} \sqrt{M^2 - 4|j_R|}\right) + O(M^0).
\end{equation}
We can write the maximum in the large $\alpha' M^2 \gg 1$ and fixed $J/M^2$ limit, which gives
\begin{equation}
    S(M,J) = \tilde\beta_{H} \sqrt{M^2 - 2J} + O(M^0),
\end{equation}
with the Hagedorn radius $\tilde R_H = \tilde \beta_H/(2\pi)$
\begin{equation}
\begin{split}
    \tilde R_{H} &= \sqrt{\frac{R_{H,L}^2+R_{H,R}^2}{2}}
    = 
    \begin{cases}
        2, & \text{Bosonic}\\
        \sqrt{2}, & \text{Type II} \\
        \sqrt{3}, & \text{Heterotic}.
    \end{cases}.
\end{split}
\end{equation}
Notice that for heterotic strings $\tilde \beta_{H} \ne \beta_H$, since for large $J$ the angular momentum is spread asymmetrically between the left and right movers, leading to a different entropy scaling.

\subsection{The \titlemath{$J/M$} expansion}
\label{app:J/M free string}
Here, we derive some formulas for the free string in the scaling $M^2 \gg 1$, with fixed $J/M\ll 1$, following \cite{Russo:1994ev}. The starting point is the same as before: the worldsheet canonical partition function $Z(b,\nu)$. We focus first on the left chirality, and then combine it with the right chirality. As we will see, in terms of the worldsheet parameters, we take the limit of constant $\nu/b$ and small $b$. In the language of~\cite{Russo:1994ev}, this corresponds to the case of constant $\lambda:= 2\pi \nu/b$. The asymptotic level density for bosonic string theory in that case is~\cite{Russo:1994ev}
\begin{equation}\label{eq:z_b_nu}
    Z(b,\nu) \sim b^{\frac{26-2}{2}} e^{\frac{\beta_H^2}{4b}}\frac{\frac{\nu}{b}}{\sinh (\frac{2\pi^2\nu}{b})},
\end{equation}
where we work in $26$ spacetime dimensions. Using a Fourier transform of \eqref{eq:z_b_nu}, we can find the  asymptotic formula for a given left-moving spin $j_L$ and occupation number $N_L$
\begin{equation}\label{bosonic level density j,N}
    Z(N_L,j_L)\sim (N_L+1-|j_L|)^{-\frac{24+5}{4}} \exp [\frac{\beta_{H,L}(N_L-\frac{|j_L|}{2})}{\sqrt{N_L+1 - |j_L|}}] \frac{1}{\cosh^2 (\frac{\beta_{H,L} j_L}{4\sqrt{N_L+1-|j_L|}})},
\end{equation}
where $\beta_{H,L}=4\pi$ is the contribution of the left chirality to the bosonic ($\nu=0$) Hagedorn temperature. The power of the prefactor has two contributions: the $24=26-2$, which is the bosonic central charge in light-cone quantization, and the (theory-independent) $5$, which comes from the Cardy formula and the $J$-measure \cite{Russo:1994ev}. 
Expanding \eqref{bosonic level density j,N} in $j_L/\sqrt{N_L}$ gives the asymptotic density of states
\begin{equation}
    Z(N_L,j_L)\sim (N_L)^{-\frac{26+3}{4}} \exp \left(\beta_{H,L}\sqrt{N_L} -  \frac{\beta_{H,L}^2}{8} \frac{j_L^2}{N_L}\right).
\end{equation}
It is straightforward to write down the asymptotic density for a  $\mathcal{N}=2$ SUSY chirality~\cite{Russo:1994ev}
\begin{equation}\label{superstring micro state}
    Z(N_L,j_L) \sim (N_L)^{-\frac{12+5}{4}}\exp\left(\beta_{H,L}\sqrt{N_L}-\frac{\beta_{H,L}^2}{8}\frac{j_L^2}{N_L}\right),
\end{equation}
where now $\beta_{H,L} = 2\sqrt{2}\pi$, and the (SUSY)  light-cone central charge in \eqref{superstring micro state} is changed to $12$.

Combining the two chiralities, level matching dictates that $N_L\approx N_R=N/2$, and so we have the formula 
\begin{equation}
    Z(N,j_L,j_R)\sim N^{-\frac{\alpha_1+2}{2}} \exp\left((\beta_L+\beta_R)\sqrt{\frac{N}{2}} - \frac{\beta_L^2}{4}\frac{j_L^2}{N}- \frac{\beta_R^2}{4}\frac{j_R^2}{N}\right),
\end{equation}
with 
\begin{equation}
    \alpha_1 = \begin{cases}
        27, & \text{Bosonic}\\
        15, & \text{Type II}\\
        21,& \text{Heterotic}
    \end{cases}.
\end{equation}
Going to fixed $J=j_L+j_R$ via a saddle point approximation gives
\begin{equation}
    Z(N,J)\sim  N^{-\frac{\alpha_1 + 1}{2}} \exp \left((\beta_L+\beta_R)\sqrt{\frac{N}{2}} - \frac{\beta_L^2\beta_R^2}{\beta_L^2+\beta_R^2} \frac{J^2}{4N}\right).
\end{equation}
Finally, using the mass-shell relation $N=M^2/2$ and transforming the measure, we get the entropy
\begin{equation}
    S(M,J)= \beta_H M -\alpha_1 \log M -\alpha_2 \frac{J^2}{M^2},
\end{equation}
with $\beta_H$ the $\nu=0$ Hagedorn temperature and 
\begin{equation}
    \alpha_2 = 2(2\pi)^2 \cdot
    \begin{cases}
        \frac{1}{2}, & \text{Bosonic}\\
        \frac{1}{4}, & \text{Type II}\\
        \frac{1}{3},& \text{Heterotic}
    \end{cases}.
\end{equation}
After restoring units, this recovers \eqref{eq:entropy_J_M} and \eqref{eq:a1a2}. 

Secondly, we would like to estimate the sizes $L_{||}, L_{\perp}$ of the free rotating string in orders of $J/M$. 
We will start in the worldsheet grand canonical ensemble, and then transform the answer to a microcanonical statement. We do this for the bosonic string, as we are only interested in order-of-magnitude results.

Following~\cite{Ceplak:2024dxm}, we estimate the sizes by taking averages of the worldsheet position for different axes.
For the size orthogonal to the plane of rotation $L_{\perp}^2$, we compute
\begin{equation}
    L_{\perp}^2 = \frac{1}{l} \int_0^{l}d\sigma \langle (X^i(\tau,\sigma))^2 \rangle,
\end{equation}
where $X^i$ is one of the transverse directions and the expectation value is taken with respect to the worldsheet grand canonical ensemble. For the size in the plane of rotation, we compute 
\begin{equation}
    L_{||}^2 = \frac{1}{2l} \int_0^{l}d\sigma  \langle (X^1(\tau,\sigma))^2 + (X^2(\tau,\sigma))^2 \rangle.
\end{equation}
Due to the worldsheet decoupling of the modes, this quantity is independent of $\nu$.
From the standard mode expansion of the free string, we find
\begin{equation}
    L_{\perp}^2 = 2 \sum_{n=1}^{\infty} \frac{1}{n} \langle N_n^i \rangle = 2 \sum_{n=1}^{\infty} \frac{1}{n} \frac{1}{e^{bn}-1} = \frac{\pi^2}{3b}(1 + \mathcal{O}(b)),
\end{equation}
where in the last step we have taken a continuum limit for high temperatures (small $b$). 
For $L_{||}^2$, however, the Bose-Einstein statistic is twisted by the chemical potential:
\begin{equation}
    \begin{split}
        L_{||}^2 &=  \sum_{n=1}^{\infty} \frac{1}{n}\left(\frac{1}{e^{bn + 2\pi i \nu}-1}+\frac{1}{e^{bn - 2\pi i \nu}-1}\right) \\
        &=\frac{\pi^2}{3b}\left(1 - \frac{4\pi^4\nu^2}{15b^2} +\mathcal{O}(b)+\mathcal{O}\left(\nu^4\right)\right).
    \end{split}
\end{equation}
The perturbative expansion of slow rotation is $\nu/b \ll 1$. 

In the second step, we would like to transform the formulas for $L_{\perp}$, $L_{||}$ into a microcanonical target space statement. For that, we calculate $M^2\approx N/4$ as a function of the worldsheet temperature from \eqref{nu b partition function}:
\begin{equation}
    M^2 \approx  -4\partial_b \log Z = \frac{(4\pi)^2}{b^2}\left(1 + \mathcal{O}(b)+\mathcal{O}\left(\frac{\nu^2}{b}\right)\right)
\end{equation}
(where we have used $\beta_H=4\pi$). Inverting the relation gives
\begin{equation}
    b= \frac{4\pi}{M}\left(1 + \mathcal{O}(M^{-1})+\mathcal{O}(\nu^2 M)\right).
\end{equation}
From the partition function, we can also compute the expectation value of $J$ by taking a derivative of \eqref{nu b partition function} with respect to $\nu$, which yields (restoring units)
\begin{equation}
    \begin{split}
        J(\nu,b)& =\frac{2}{2\pi i}\partial_{\nu}\log Z(b,\nu)=2\sum_{n=1}^{\infty} \left(\frac{1}{e^{bn-2\pi i \nu}-1}-\frac{1}{e^{bn+2\pi i \nu}-1}\right) \\& =\frac{4\pi^3 i \nu}{3b^2}(1+\mathcal{O}(b))+\mathcal{O}(\nu^3) = \frac{\pi(l_s M)^2}{12} i \nu(1+ \mathcal{O}(M^{-1})) +\mathcal{O}(\nu^3 M^4),
    \end{split}
\end{equation}
where the factor of $2$ in the enumerator is from the fact that we have contributions from both chiralities. 

This also allows us to express the length in the microcanonical ensemble: 
\begin{equation}
    \begin{split}
        L_{||} & = \frac{l_s^{\frac{3}{2}} \sqrt{\pi M}}{\sqrt{12}}\left(1+\mathcal{O}(M^{-1})+\frac{12J^2}{5(l_s M )^2}+\mathcal{O}\left(\frac{J^4}{M^4}\right) \right) \\
        L_{\perp} & = \frac{l_s^{\frac{3}{2}} \sqrt{\pi M}}{\sqrt{12}}\left(1+\mathcal{O}(M^{-1})  + \mathcal{O}\left(\frac{J^2}{M}\right) +\mathcal{O}\left(\frac{J^4}{M^4}\right)\right),
    \end{split}
\end{equation}
reproducing \eqref{eq:L_J_M}. 

\section{Fixed angular velocity in Newtonian physics}
\label{app:newton}
This appendix reviews general properties of classical non-relativistic systems at fixed angular velocity. Let $\vec x_i(t)$ ($i=1,...,N$) be $N$ classical particles with masses $m_i$ and a joint potential energy $V(\vec x_j)$. We assume the potential is invariant under global translations and rotations of the positions $\vec x_i$. For brevity, we will assume two spatial dimensions, but the general lesson is easily generalized to higher dimensions.
The classical action of the system is given by
\begin{equation}\label{eq:S_L}
    S[\vec x_i] = \int dt \left(\sum_i \frac{1}{2} m_i \dot x_i^2(t)- V(\vec x_j(t))\right).
\end{equation}

Consider the partition function of the system
\begin{equation}\label{eq:Z_part}
    Z(\beta,\Omega) = \text{Tr}\left(\exp(-\beta (H-\Omega J))\right),
\end{equation}
with $J$ the angular momentum operator
\begin{equation}
    J = \sum_i \left(p^1_i x^2_i-p^2_i x^1_i\right),
\end{equation}
and $p^a_{i}$ the conjugate momenta of $x^a_i$. 
In Lagrangian variables, the partition function can be written as a Euclidean path integral, with the Euclidean action
\begin{equation}\label{eq:rot_frane_SE_1}
    S_E[\vec x_i] = \int_0^\beta d\tau \left(\sum_i \frac{m_i}{2}\left( (\dot x_i^1-i \Omega x_i^2)^2 + (\dot x_i^2+i \Omega x_i^1)^2\right)+V(\vec x_j)\right),
\end{equation}
where $0\le \tau\le \beta$ is the Euclidean time. This action is simply the (Wick-rotated) rewriting of \eqref{eq:S_L} in a rotating reference frame with angular velocity $\Omega$. This is clear as instead of time evolving by $H$, we evolve in \eqref{eq:Z_part} by $H-\Omega J$. By itself, changing the reference frame does nothing to the space of solutions. 
Due to the trace, however, the path-integral sums only over periodic paths $\vec x_i(0)=\vec x_i(\beta)$ in the rotating frame. In the classical thermodynamic limit, we further look for solutions that are stationary in $\tau$.
In terms of the original frame, this is the condition $\vec x_i(\tau) = e^{\hat J\cdot \Omega \tau } \vec x_i(0)$. Therefore, \eqref{eq:Z_part} only includes configurations with constant angular velocity $\Omega$.

Let us assume that at $\Omega=0$, the non-rotating system possesses a (stable) stationary solution $\vec x_i(t) = \vec r_i$, with $\partial V/\partial \vec x_i(\vec r_j) = 0$. We further assume that the center of mass is at rest in $\vec x_{CM}=0$.
In the Euclidean action \eqref{eq:rot_frane_SE_1}, we can expand in order of $\Omega$ using perturbation theory. To leading order in perturbation theory, the action is given by substituting the non-rotating solution:
\begin{equation}
    S_E = \beta \cdot \left(V(\vec r_i) - \frac{I }{2} \Omega^2 + O(\Omega^4)\right),
\end{equation}
with the moment of inertia
\begin{equation}
    I = \sum_i m_i r_i^2.
\end{equation}
The first term is the rest energy, and the second term is the standard rigid-body expression.
In terms of the non-rotating reference frame, we found that to order $O(\Omega^2)$, the constant angular velocity solution is simply the rotation of the $\Omega=0$ solution $\vec x_i(t) = e^{\Omega \tau \hat J} \vec r_i$. This is indeed the rigid-body approximation, which is expected to hold as long as the centrifugal force is parametrically smaller than the internal forces of the system. At this order, the averaged angular momentum of the system is given by the rigid-body answer
\begin{equation}
    \langle J \rangle = \frac{\partial}{\partial \Omega} T \log Z = I \cdot \Omega + O(\Omega^3).
\end{equation}

To study the system more generally, we can assume a stationary solution (in the rotating frame) $\vec x_i$ to get for the Euclidean action
\begin{equation}\label{eq:rot_frane_SE_2}
    S_E(\vec x_i) = \beta \left(V(\vec x_j)-\frac{1}{2} \Omega^2 \sum_i m_i x_i^2\right).
\end{equation}
The second term is simply the potential for the centrifugal force. Classical solutions satisfy $\partial S_E/\partial \vec x_j=0$ - a balance between the internal forces $V'(x)$ and the centrifugal force. The leading correction to the non-rotating solution would (generically) be of order $\delta \vec x_i \sim \Omega^2$ to cancel the centrifugal force. This correction will contribute to the action (or the energy) at order $O(\Omega^4)$.

For a generic potential, we can always expect a classical stationary solution for some finite range of $\Omega$s close enough to zero.
However, since generic potentials decay as a power law or faster, the potential in \eqref{eq:rot_frane_SE_2} is always unbounded from below for large $x^2_i$s. For this reason, the classical solution we found for $\Omega>0$ is always meta-stable in this ensemble.

\bibliographystyle{JHEP}
\bibliography{ref}
\end{document}